\documentclass[aps,prd,twocolumn,nofootinbib,showpacs,amsmath,amssymb,floatfix,eqsecnum]{revtex4-1}

\usepackage{graphicx,color,framed}
\usepackage{hyperref}
\usepackage{times}
\usepackage{enumerate}
\usepackage{lipsum}
\usepackage{slashed}
\usepackage{array}
\usepackage{textcomp}

\hypersetup{
    colorlinks=true, 
    linktoc=all,     
    linkcolor=blue,  
}

\def \beq {\begin{equation}}
\def \eeq {\end{equation}}
\def \beqa {\begin{eqnarray}}
\def \eeqa {\end{eqnarray}}
\def \bseq {\begin{subequations}}
\def \eseq {\end{subequations}}
\newcommand \dg {\dagger}

\newcommand \al {\alpha}

\newcommand \ran {\rangle}
\newcommand \lan {\langle}

\newcommand \ep {\epsilon}
\newcommand \lam {\lambda}
\newcommand \pd {\partial}

\newcommand \mb {\mathbf}
\newcommand \mbs {\boldsymbol}
\newcommand \nnb {\nonumber}

\newcommand \M {\mathcal{M}}
\newcommand \ov {\overline}
\newcommand \td {\tilde}

\begin{document}

\title{On the parity anomaly from the Hamiltonian point of view}

\author{Matthew F. Lapa}
\email[email address: ]{mlapa@uchicago.edu}
\affiliation{Kadanoff Center for Theoretical Physics, University of Chicago, Illinois 60637, USA}


\begin{abstract}

We review the parity anomaly of the massless Dirac fermion in $2+1$ dimensions from the Hamiltonian, as opposed to the path integral,
point of view. We have two main goals for this note. First, we hope to make the parity anomaly more accessible to 
condensed matter physicists, who generally prefer to work within the Hamiltonian formalism. The parity anomaly plays an important role in 
modern condensed matter physics, as the massless Dirac fermion is the surface theory of the time-reversal invariant topological insulator (TI) in 
$3+1$ dimensions. Our second goal is to clarify the relation between the time-reversal symmetry of the massless Dirac fermion and the fractional 
charge of $\pm\frac{1}{2}$ (in units of $e$) that appears on the
surface of the TI when a magnetic monopole is present in the bulk. To accomplish these goals we study the Dirac fermion in the Hamiltonian 
formalism using two different regularization schemes. One scheme is consistent with the time-reversal symmetry of the massless Dirac fermion, 
but leads to the aforementioned fractional charge. The second scheme does not lead to any fractionalization, but it does break 
time-reversal symmetry. For both regularization schemes we also compute the effective action $S_{\text{eff}}[A]$ that encodes the response of 
the Dirac fermion to a background electromagnetic field $A$. We find that the two effective actions differ by a Chern-Simons counterterm
with fractional level equal to $\frac{1}{2}$, as is expected from path integral treatments of the parity anomaly. Finally, we propose the
study of a bosonic analogue of the parity anomaly as a topic for future work.

\end{abstract}


\maketitle

\section{Introduction}

The purpose of this note is to review the \emph{parity anomaly} of the
massless Dirac fermion in $2+1$ dimensions~\cite{NS,redlich1,redlich2,anomalies-odd}, 
but from the Hamiltonian/Hilbert space point of view. Recall that the parity anomaly is a conflict between the 
time-reversal\footnote{As emphasized by Witten~\cite{witten-parity}, 
the word ``parity" in ``parity 
anomaly" is a misnomer, and this anomaly is actually an anomaly in time-reversal or reflection symmetry.} symmetry
and large $U(1)$ gauge invariance of the massless Dirac fermion. 
More precisely, the parity anomaly is equivalent to the statement that it is 
impossible to regularize the massless Dirac fermion theory, coupled to a background $U(1)$ gauge field, in a way that preserves both time-reversal 
symmetry and large $U(1)$ gauge invariance. To clarify the meaning of large $U(1)$ gauge invariance here, note that unbroken large
$U(1)$ gauge invariance would require all physical states of the theory to have integer charge, and so any regularization that leads to states with 
fractional charge must violate large $U(1)$ gauge invariance.

There are two main reasons why we
feel that a review of the parity anomaly from the Hamiltonian perspective is warranted. 
First, the parity anomaly has been discussed extensively in recent years in the context of the time-reversal
invariant topological insulator (TI)~\cite{fu-kane-mele,QHZ}, which hosts a single massless Dirac fermion on its surface. In this
context the parity anomaly provides one of the classic examples of a theory with
a \emph{\textquotesingle t Hooft} anomaly~\cite{hooft1980naturalness} appearing at the boundary of a 
symmetry-protected topological phase~\cite{wen2013classifying,kapustin2014symmetry,kapustin2014anomalies,kapustin2014anomalous}. 
However, the discussion in the recent literature on this topic is almost\footnote{One exception is the
recent mathematical treatment of the parity anomaly in Ref.~\onlinecite{muller-szabo}. There the authors studied the 
projective representation of the $U(1)$ gauge group on the Hilbert space of the massless Dirac fermion.} 
always from the path integral point of 
view~\cite{mulligan-burnell,witten-fermions,seiberg-witten,witten-parity,PhysRevD.96.025011,kurkov2018gravitational}. 
On the other hand, in condensed matter physics it is more common to look at problems from a 
Hamiltonian point of view. Therefore we believe that there is significant value in explaining how the parity anomaly works from the point of view of 
the Dirac Hamiltonian on two-dimensional (2D) \emph{space}. It is also worth noting that the
TI is one of the few symmetry-protected topological phases that have been realized experimentally (see \cite{hsieh2008topological} and the review 
\cite{hasan-kane}), and so further study and clarification of the parity anomaly in the context of TI physics seems justified.

The second reason for our review of the parity anomaly is to explain the precise connection between the
time-reversal symmetry of the massless Dirac fermion and the half-quantized electric charge of $\pm\frac{1}{2}$ (in units of $e$)
that appears on the surface of the TI when a magnetic monopole is present in the bulk. In Ref.~\onlinecite{NS}, Niemi and Semenoff
studied the \emph{massive} Dirac fermion in the Hamiltonian formalism using a regularization scheme based on the 
Atiyah-Patodi-Singer (APS) \emph{eta invariant}~\cite{APS1}, or \emph{spectral asymmetry} of the Dirac Hamiltonian on 2D \emph{space}.
Within this scheme they found that the ground state of the massive Dirac fermion has a charge of $\pm \frac{1}{2}$ when the 2D space is pierced 
by a single unit of magnetic flux, and they also found that this charge persists in the limit in which the mass of the fermion is sent to zero.
The fractional charge of 2D electrons in a magnetic field was also studied by Jackiw in Ref.~\onlinecite{jackiw1984}.

In this note we point out that the regularization scheme used by Niemi and Semenoff is consistent with the time-reversal symmetry of the 
massless Dirac fermion, in a sense that we make precise below. To the best of our knowledge, the fact that the regularization scheme in 
\cite{NS} is consistent with time-reversal symmetry has not been demonstrated in detail in the existing literature. 
The closest discussion that we know of can be found in Ref.~\onlinecite{barci2001point}, where it was shown that the results obtained
by Niemi and Semenoff are identical to the results obtained from a \emph{point-splitting} regularization scheme that preserves parity (and
time-reversal) symmetry. 
The fact that the regularization in \cite{NS} is consistent with time-reversal should not be unexpected though, as it fits in with 
the general picture of the parity anomaly discussed above (i.e., the 
regularization of \cite{NS} violates large $U(1)$ gauge invariance, so we expect that it should be consistent with time-reversal symmetry). 
This fact also makes the regularization scheme of \cite{NS} the correct scheme to use in the physical situation where the massless Dirac fermion 
resides on the surface of the TI.

In the path integral approach, which was pioneered by Redlich~\cite{redlich1,redlich2}, the easiest way to see the parity anomaly is to use 
\emph{Pauli-Villars} regularization to compute the partition function of the massless Dirac fermion. 
This regularization scheme preserves large $U(1)$ gauge invariance, but it breaks time-reversal symmetry because of the mass of the 
Pauli-Villars regulator fermion. An alternative regularization scheme, also considered by Redlich, is to define the partition function of the massless 
Dirac fermion as the square root of the determinant of a Dirac operator for two copies of a massless Dirac fermion. This
latter determinant can be regularized in a time-reversal invariant way, which leads to a time-reversal invariant regularization of the 
original single massless Dirac fermion. Redlich then showed that this second regularization scheme violates large
$U(1)$ gauge invariance. We also note here that in a more sophisticated 
treatment~\cite{anomalies-odd,witten-fermions,seiberg-witten,witten-parity},
Pauli-Villars regularization leads to an expression for the partition function of the massless Dirac fermion in which the phase
of the partition function is proportional to the APS eta invariant of the \emph{spacetime} Dirac operator. The
APS eta invariant is constructed from the spectrum of the Dirac operator, and so it is manifestly gauge invariant, but this scheme still breaks
time-reversal symmetry, again due to the mass of the regulator fermion.

The purpose of this note is to explain how to see the conflict between time-reversal symmetry and large U(1) gauge invariance when the
massless Dirac fermion is studied from the Hamiltonian point of view. To this end, we study the Dirac fermion in the Hamiltonian
formalism using two different regularization schemes. The first regularization scheme leads
to states in the theory with half-integer charge, but we show that this scheme is consistent with time-reversal symmetry. 
The second regularization scheme explicitly breaks time-reversal symmetry but does not lead to any fractionalized quantum numbers
associated with the $U(1)$ symmetry. Thus, these two regularization schemes serve to demonstrate the parity anomaly in the Hamiltonian/Hilbert 
space approach.

The first regularization scheme that we consider is exactly the scheme used by Niemi and Semenoff~\cite{NS}. 
For this scheme we work on a general curved two-dimensional space $\M$ that is a closed\footnote{We consider closed manifolds (e.g., 
the two-sphere $S^2$) instead of $\mathbb{R}^2$ to make the problem mathematically simpler. In particular, on closed manifolds the
Dirac operator has discrete eigenvalues, and in this case we can also apply the Atiyah-Singer index theorem to answer certain questions
regarding the zero modes of the Dirac operator. See Ref.~\onlinecite{jackiw1984} for a discussion of the difference between the case of the
plane $\mathbb{R}^2$ and the case of closed manifolds.} manifold, instead of on 
flat space $\mathbb{R}^2$. The specific physical quantity that we calculate within this regularization scheme is $Q_{A,m}$, 
the charge (in units of $e$) 
of the ground state of the theory in the presence of a time-reversal breaking mass term (with mass $m$), and in the presence of a background
time-independent spatial gauge field $A=A_j dx^j$ (we use differential form notation and also sum over the spatial index $j=1,2$).  
We show that the regularization scheme of \cite{NS} is consistent with time-reversal symmetry in the sense that it leads to the result
\beq
	Q_{A,m}=Q_{-A,-m}\ . \label{eq:GS-charge-TR}
\eeq
Physically, this result means that in this regularization scheme the charge in the ground state of the theory with mass $m$ and background field 
$A$ is equal to the charge in the ground state of the \emph{time-reversed} theory with mass $-m$ and background field $-A$ (a spatial gauge field is 
odd under time-reversal). On the other hand,
the explicit result for $Q_{A,m}$ (Eq.~\eqref{eq:ground-q} in Sec.~\ref{sec:reg1}) shows that it can be integer or half-integer valued,
\beq
	Q_{A,m} \in \frac{1}{2}\mathbb{Z}\ ,
\eeq
which shows that this regularization scheme violates large $U(1)$ gauge invariance.

The second regularization scheme that we consider is a \emph{lattice} regularization scheme
for the massless Dirac fermion on a spatial torus. The lattice model that we use for this regularization is
on the square lattice, but this model is closely related to the model on the honeycomb lattice that was introduced in the 
seminal work of Haldane~\cite{haldane-parity} on a model for the quantum Hall effect without Landau levels.
The specific physical quantity that we calculate in this scheme is $\sigma_{H,m}$, the Hall conductivity 
(in units of $\frac{e^2}{h}$) of the Dirac fermion with mass $m$ in the 
presence of a background time-independent electric field $\mb{E}$. We find that $\sigma_{H,m}$ is given by
\beq
	\sigma_{H,m}= \frac{\text{sgn}(m)-1}{2} \in \mathbb{Z}\ .
\eeq
This result demonstrates two things. First, the Hall conductivity is an integer for either sign of $m$, which shows that  
large $U(1)$ gauge invariance is preserved by this regularization scheme (there is no fractionalization of quantum numbers associated with the
$U(1)$ symmetry). Second, the Hall conductivity for the theory with mass $m$ is 
\emph{not} equal to minus the Hall conductivity for the time-reversed theory with mass $-m$,
\beq
	\sigma_{H,m} \neq -\sigma_{H,-m}\ .
\eeq
This shows that this regularization scheme is not consistent with the time-reversal symmetry of the original massless Dirac fermion (a 
regularization scheme consistent with time-reversal symmetry should give $\sigma_{H,m} = -\sigma_{H,-m}$ since the Hall conductivity is
odd under time-reversal). 

Note that in both cases we never treat the massless theory directly --- the quantities $Q_{A,m}$ and 
$\sigma_{H,m}$ that we study are both computed for the theory with a non-zero time-reversal breaking mass $m$. 
Instead, we determine whether the result is consistent with time-reversal symmetry by comparing the answers for
two massive theories that are related to each other by the time-reversal operation.

Finally, for both regularization schemes we also compute the effective action $S_{\text{eff}}[A]$ that encodes the response of the massive theory
to the background gauge field $A=A_{\mu}dx^{\mu}$ ($\mu=0,1,2$). 
We find that the effective action $S^{\text{(NS)}}_{\text{eff}}[A]$, computed using the regularization scheme of Niemi and Semenoff, is related to the 
effective action $S^{\text{(lattice)}}_{\text{eff}}[A]$, computed using the lattice regularization, as
\beq
	S^{\text{(lattice)}}_{\text{eff}}[A]= S^{\text{(NS)}}_{\text{eff}}[A] -\frac{1}{2}\frac{1}{4\pi}\int A\wedge dA\ . \label{eq:CS-counterterm}
\eeq
The last term on the right-hand side is a \emph{Chern-Simons} term (written in differential form notation), but with a \emph{fractional level} 
equal to $-\frac{1}{2}$. Thus, the two effective actions differ by a Chern-Simons counterterm with fractional level $-\frac{1}{2}$, which is
exactly the result that we expect based on the original path integral treatment of the parity anomaly~\cite{redlich1,redlich2}.

This note is organized as follows. In Sec.~\ref{sec:Dirac-fermion} we review the form of the Hamiltonian for the Dirac fermion on flat and curved 
2D space, and we also review the time-reversal symmetry of the massless Dirac fermion. In Sec.~\ref{sec:reg1} we study the Dirac fermion on a 
closed spatial manifold $\M$ using the regularization scheme of Niemi and Semenoff~\cite{NS}, 
and we compute the charge $Q_{A,m}$ of the ground state for the
massive Dirac fermion in the presence of a background time-independent spatial gauge field $A$. 
In Sec.~\ref{sec:reg2} we study the Dirac fermion using a 
lattice regularization scheme on a spatial torus, and we compute the Hall conductivity $\sigma_{H,m}$ for the massive Dirac fermion in the
presence of a time-independent electric field $\mb{E}$. In Sec.~\ref{sec:effective-action} we compute the effective action $S_{\text{eff}}[A]$ for 
both regularization schemes, and we show that the two effective actions are related as shown in Eq.~\eqref{eq:CS-counterterm}. 
In Sec.~\ref{sec:con} we 
present concluding remarks and propose the study of a similar anomaly in bosonic systems for future work. 
Finally, Appendix~\ref{app:Dirac} contains important background
material on Dirac fermions on curved space and on the notation used in the paper. 

\textbf{Note:} Throughout the paper we work in system of units where the Dirac fermion has charge $e=1$, and where $\hbar=1$ (so 
$h=2\pi\hbar \to 2\pi$) and $c=1$.
Here $c$ would be the speed of light in a high-energy context or the Fermi velocity in a condensed matter context. We use a summation
convention in which we sum over any index that appears once as a subscript \emph{and} once as a superscript in any expression, and 
we use Latin indices $j,k,\dots$ taking values $\{1,2\}$ to label spatial directions and Greek indices $\mu,\nu,\dots$ taking values $\{0,1,2\}$ to 
label spacetime directions. We also use Latin indices $a,b,\dots$ near the beginning of the alphabet for frame indices on curved space 
(see Appendix~\ref{app:Dirac}).
In general, we recommend that readers glance at Appendix~\ref{app:Dirac} before reading the paper, to make sure that they are familiar with our 
notation and conventions for the Dirac operator on curved space, and also to review the relation between the $U(1)$ gauge field 
$A=A_{\mu}dx^{\mu}$ and the ordinary electric and magnetic fields $\mb{E}$ and $B$ on flat space.

\section{Dirac Hamiltonian and time-reversal symmetry}
\label{sec:Dirac-fermion}

In this section we introduce the Dirac fermion on flat and curved two-dimensional \emph{space}. We also discuss the time-reversal symmetry
of the massless Dirac fermion, and we discuss the effect of time-reversal on the Dirac fermion with non-zero mass $m$ and in the presence of 
a background time-independent spatial $U(1)$ gauge field $A=A_j dx^j$. 

\subsection{Flat space}

We start with the action for the massless Dirac fermion on flat Minkowski spacetime, 
\beq
	S[\Psi,\ov{\Psi}]= \int d^3x\  \ov{\Psi}i\td{\gamma}^{\mu}\pd_{\mu}\Psi\ .
\eeq
The quantities appearing here are as follows.
First, $x=(x^0,x^1,x^2)$ is the spacetime coordinate, $\pd_{\mu}\equiv\frac{\pd}{\pd x^{\mu}}$ for
$\mu=0,1,2$, and $\Psi=\Psi(x)$ is a two-component Dirac spinor field on spacetime. Next, $\td{\gamma}^{\mu}$ are a set of gamma matrices
that satisfy the Clifford algebra $\{\td{\gamma}^{\mu},\td{\gamma}^{\nu}\}= 2\eta^{\mu\nu}$, where 
$\eta= \text{diag}(1,-1,-1)$ is the Minkowski metric in ``mostly minus'' convention. Finally, $\ov{\Psi}= \Psi^{\dg}\td{\gamma}^0$ is the
Dirac adjoint of $\Psi$.

Next, we discuss the coupling to a background $U(1)$ gauge field (electromagnetic field) represented by the vector potential one-form
$A=A_{\mu}dx^{\mu}$. Our convention is that the Dirac fermion has charge $1$. The correct action for $\Psi$ coupled to $A$ is then
\beq
	S[\Psi,\ov{\Psi},A]= \int d^3x\  \ov{\Psi}i\td{\gamma}^{\mu}(\pd_{\mu}+i A_{\mu})\Psi\ .
\eeq
To see that the sign of the coupling to $A_{\mu}$ is correct for fermions with charge $1$, note that the term with $A_0$ is 
\beq
	-\int d^3x\  \Psi^{\dg}\Psi\ A_0\ ,
\eeq
and this is the correct action for a distribution of charge with charge density $\Psi^{\dg}\Psi$ in the presence of a scalar electromagnetic 
potential $A_0$ (for charge $e$ the correct covariant derivative is $\pd_{\mu}+i e A_{\mu}$). We refer the reader to the end of 
Appendix~\ref{app:Dirac} for more details on how the components of the one-form $A$ are related to the usual
electric and magnetic fields $\mb{E}$ and $B$ in the case of flat Minkowski spacetime. 

Finally, the mass term for the Dirac fermion takes the simple form
\beq
	S_m[\Psi,\ov{\Psi}]= -m\int d^3x\  \ov{\Psi}\Psi\ ,
\eeq
where the mass $m$ is a real parameter that can be positive or negative.

We now pass to the Hamiltonian formulation of the massless Dirac fermion on flat space. The momentum canonically conjugate to
$\Psi$ is $i\Psi^{\dg}$. As a result, the Dirac Hamiltonian on flat 2D space takes the form
\beq
	\hat{H}=  -\int d^2 \mb{x}\ \hat{\Psi}^{\dg} i\td{\gamma}^0\td{\gamma}^j\pd_j\hat{\Psi}\ ,
\eeq
where $\mb{x}=(x^1,x^2)$ is the spatial coordinate, $j=1,2$, and $\hat{\Psi}= \hat{\Psi}(\mb{x})$ is the operator-valued Dirac spinor on 
2D space.\footnote{Later on we define the operator $\hat{\Psi}(\mb{x})$ 
more precisely using a mode expansion in terms of eigenfunctions of the
appropriate Dirac differential operator on flat or curved space --- see Eq.~\eqref{eq:mode-expansion}.}
To proceed, it is convenient to define a new set of spatial gamma matrices by 
$\gamma^j= -\td{\gamma}^0\td{\gamma}^j$. These
new gamma matrices obey the Clifford algebra $\{\gamma^j,\gamma^k\}=2\delta^{jk}$. In addition, we define the Dirac (differential) operator 
$\mathcal{H}$ on 2D space by
\beq
	\mathcal{H}= i\gamma^j\pd_j\ .
\eeq
In terms of these new quantities the massless Dirac Hamiltonian on flat space takes the form
\beq
	\hat{H} =  \int d^2 \mb{x}\ \hat{\Psi}^{\dg}\mathcal{H}\hat{\Psi}\ .
\eeq

The Hamiltonian $\hat{H}$ for the massless Dirac fermion commutes with a time-reversal operator $\hat{T}$ that is defined as follows. 
First, we define a third gamma matrix $\overline{\gamma}= \frac{i}{2}\ep_{jk}\gamma^{j}\gamma^{k}= i\gamma^1\gamma^2$, which
satisfies $\{\ov{\gamma},\gamma^j\}=0$ and $\ov{\gamma}^2=1$. The matrix $\ov{\gamma}$ is sometimes referred to as the
\emph{chirality} matrix. Next, we choose a concrete realization for the three gamma matrices
$\gamma^j$ ($j=1,2$) and $\ov{\gamma}$ such that the $\gamma^j$ have real matrix elements and $\ov{\gamma}$ has \emph{imaginary}
matrix elements. For example we could choose $\gamma^1=\sigma^x$, $\gamma^2=\sigma^z$, and then $\ov{\gamma}=\sigma^y$, where
$\sigma^{x,y,z}$ are the Pauli matrices. 

With these conventions in place, the action of the time-reversal operator $\hat{T}$ on $\hat{\Psi}$ is defined to be
\begin{subequations}
\label{eq:TR-def}
\beqa
	\hat{T} \hat{\Psi}_{\al} \hat{T}^{-1} &=& {\ov{\gamma}_{\al}}^{\beta}\hat{\Psi}_{\beta} \\
	\hat{T} \hat{\Psi}^{\dg,\al} \hat{T}^{-1} &=& \hat{\Psi}^{\dg,\beta}{\ov{\gamma}_{\beta}}^{\al}\ ,
\eeqa 
\end{subequations}
where $\hat{\Psi}_{\al}$, $\al=1,2$, are the two components of the spinor-valued field $\hat{\Psi}$, $\hat{\Psi}^{\dg,\al}$ are
the two components of $\hat{\Psi}^{\dg}$, and ${\ov{\gamma}_{\al}}^{\beta}$ are the matrix elements of $\ov{\gamma}$. As usual, 
$\hat{T}$ is an \emph{anti-unitary} operator, so it will complex-conjugate any c-numbers that it passes through. With this definition
of $\hat{T}$ we find that $\hat{T}^2  \hat{\Psi}_{\al}\hat{T}^{-2}= -\hat{\Psi}_{\al}$ and likewise for 
$\hat{\Psi}^{\dg}$ (this property is usually summarized by the equation $\hat{T}^2= (-1)^{\hat{N}}$, where $\hat{N}$ is the fermion
number operator). In addition, one can show that the massless Dirac Hamiltonian above commutes with this time-reversal 
operator
\beq
	\hat{T}\hat{H} \hat{T}^{-1}= \hat{H}\ ,
\eeq
and to show this it is necessary to use the fact that $\hat{T}$ is anti-unitary. We emphasize here that in the definition of $\hat{T}$ it was
crucial that we chose the gamma matrices so that $\ov{\gamma}$ has imaginary matrix elements and the $\gamma^j$ have real matrix elements.

In Sec.~\ref{sec:reg1} we will be interested in coupling this theory to a \emph{time-independent} background electromagnetic field which is
specified by the spatial vector potential $A=A_j dx^j$ (we do not turn on 
a time-component $A_0$ for this discussion). We will also be interested in adding a mass term to $\hat{H}$. Starting from the Dirac action
coupled to $A$ and with a non-zero mass term, it is straightforward to see that the resulting Hamiltonian takes the form
\beq
	\hat{H}_{A,m}= \int d^2 \mb{x}\ \hat{\Psi}^{\dg}\mathcal{H}_{A,m} \hat{\Psi}\ .
\eeq
where $\mathcal{H}_{A,m}$ is the massive Dirac operator coupled to $A$ on 2D space, 
\beq
	\mathcal{H}_{A,m}= i\gamma^j(\pd_j+i A_j) + m \overline{\gamma}\ .
\eeq
To arrive at this form of $\mathcal{H}_{A,m}$ we have also chosen our gamma matrices so that 
$\td{\gamma}^0=\ov{\gamma}$, where $\td{\gamma}^0$ was the 
original gamma matrix associated with the time direction. 
Since $A$ is a background field (as opposed to a quantum operator), and since it is real-valued, it commutes with $\hat{T}$.
Then we find that under time-reversal the Hamiltonian for the massive theory coupled to $A$ transforms as
\beq
	\hat{T}\hat{H}_{A,m}\hat{T}^{-1}= \hat{H}_{-A,-m}\ .
\eeq
In other words, the theories with $(A,m)$ and $(-A,-m)$ are time-reverses of each other, and only the theory with $A=0$ and $m=0$ is 
invariant under the action of $\hat{T}$.

\subsection{Generalization to curved space}

We now discuss the form of the Dirac Hamiltonian on curved space. In this case the flat two-dimensional plane $\mathbb{R}^2$ (the
spatial part of Minkowski spacetime) is replaced by a curved manifold $\M$. We assume that $\M$ is a 2D orientable Riemannian manifold.
We also assume that $\M$ is \emph{closed} (i.e., compact and without boundary) and connected. In a coordinate patch on $\M$ with coordinates
$\mb{x}=(x^1,x^2)$, the components of the metric $g$ will be denoted by $g_{jk}(\mb{x})$, and $\text{det}[g(\mb{x})]>0$
is the determinant of $g$ at the point $\mb{x}$. Since $\M$ is 2D it is also a spin manifold, and so we do not need to worry about the issue of 
whether or not fermions can be consistently placed on $\M$. 

The Hamiltonian for the massless Dirac fermion on $\M$ takes the form
\beq
	\hat{H}= \int d^2 \mb{x}\ \sqrt{\text{det}[g(\mb{x})]} \hat{\Psi}^{\dg}\mathcal{H}\hat{\Psi}\ ,
\eeq
where
\beq
	\mathcal{H}= i\slashed{\nabla}
\eeq
is the Dirac operator on $\mathcal{M}$. In Appendix~\ref{app:Dirac} we review the form of the Dirac operator on a 
general spin manifold $\M$, including our conventions for gamma matrices and so on, and we suggest that readers take a look at that 
appendix before reading the rest of this note. 

In 2D the Dirac operator simplifies greatly and we have
\beq
	\slashed{\nabla} = e^{j}_a \gamma^a \left(\pd_{j} -\frac{i}{2} \omega_{j}\overline{\gamma} \right)\ ,
\eeq
where $\gamma^a$, $a=1,2$, are gamma matrices with \emph{frame} indices, $e^{j}_a$ are the components of the 
frame vector field $e_a=e^j_a \pd_j$ on $\M$, $\omega_j$ are the components of the spin 
connection one-form $\omega=\omega_j dx^j$ on $\M$, and the matrix $\ov{\gamma}$ is now defined using the gamma matrices
with frame indices as $\overline{\gamma}= \frac{i}{2}\ep_{ab}\gamma^{a}\gamma^{b}$ (it still satisfies
$\{\ov{\gamma},\gamma^a\}=0$ and $\ov{\gamma}^2=1$). 
We can write the Dirac operator in this simplified form in 2D because in this dimension the only non-zero components of the spin connection 
${{\omega_{j}}^a}_b$ on $\M$ are ${{\omega_{j}}^1}_2 = -{{\omega_{j}}^2}_1$, and so we can write everything in terms of the single 
quantity $\omega_j := {{\omega_{j}}^1}_2$.  

The massless Dirac Hamiltonian on the curved space $\M$ has the same time-reversal symmetry as on flat space. If we choose the
gamma matrices with frame indices so that the $\gamma^a$ are real, then we again find that $\ov{\gamma}$ is imaginary,
and the time-reversal operation for the case of curved space can be defined using $\ov{\gamma}$ just as in Eq.~\eqref{eq:TR-def} on flat 
space. With that definition we again find that $\hat{T}\hat{H}\hat{T}^{-1}=\hat{H}$, so that the massless Dirac fermion is still time-reversal
invariant even on curved space.

Finally, on curved space the Hamiltonian for the massive Dirac fermion coupled to the time-independent spatial gauge field $A=A_j dx^j$ 
takes the form
\beq
	\hat{H}_{A,m}= \int d^2 \mb{x}\ \sqrt{\text{det}[g(\mb{x})]} \hat{\Psi}^{\dg} \mathcal{H}_{A,m} \hat{\Psi}\ ,
\eeq
where
\beq
	\mathcal{H}_{A,m}= i\slashed{\nabla}_A + m \overline{\gamma}\ 
\eeq
and
\beq
	\slashed{\nabla}_A = e^{j}_a \gamma^a \left(\pd_{j} + i  A_{j} -\frac{i}{2} \omega_{j}\overline{\gamma} \right)\  \label{eq:Dirac-A}
\eeq
is the massless Dirac operator on curved space and coupled to $A$.
We again find that $\hat{H}_{A,m}$ transforms under time-reversal as $\hat{T}\hat{H}_{A,m}\hat{T}^{-1}= \hat{H}_{-A,-m}$.

\section{Regularization scheme 1}
\label{sec:reg1}

In this section we study the Dirac fermion using our first regularization scheme, which is the scheme used by Niemi and 
Semenoff in Ref.~\onlinecite{NS}. In this
regularization scheme we compute the charge $Q_{A,m}$ in the ground state of the massive Dirac fermion theory on the curved space $\M$
and in the presence of the time-independent background spatial gauge field $A=A_j dx^j$. We then explain that this regularization scheme is
consistent with the time-reversal symmetry of the massless Dirac fermion, in the sense that Eq.~\eqref{eq:GS-charge-TR} holds, i.e., in the
sense that this regularization leads to equal ground state charges for the theory with $(A,m)$ and the time-reversed theory with $(-A,-m)$.

We start by introducing the \emph{normal-ordered} charge operator $\hat{Q}$ for the Dirac fermion 
\beq
	\hat{Q}= \frac{1}{2} \int d^2 \mb{x} \ \sqrt{\text{det}[g(\mb{x})]} \left[ \hat{\Psi}^{\al,\dg}(\mb{x}),\hat{\Psi}_{\al}(\mb{x}) \right]\ . \label{eq:NN-charge}
\eeq
For comparison, the non-normal-ordered version of this operator would just be the familiar expression
$\int d^2\mb{x}\ \sqrt{\text{det}[g(\mb{x})]} \hat{\Psi}^{\dg}(\mb{x})\hat{\Psi}(\mb{x})$. The reason that we use the normal-ordered charge 
operator is that the expectation value of this operator is zero in the ground state of the theory with the background field $A$ set to zero. 
Using the time-reversal operation defined above, it is simple to show that this operator is time-reversal invariant,
\beq
	\hat{T}\hat{Q}\hat{T}^{-1}=\hat{Q}\ ,
\eeq
which is exactly what we expect for the physical electric charge.

\subsection{Ground state charge and the eta invariant of the spatial Dirac operator}
\label{sec:ground-charge}

We now calculate the charge of the ground state of the massive Dirac fermion theory in the presence of the background field $A$.
Our discussion is similar to the original derivation in \cite{NS}, but adapted to the case of 
curved space. The key idea of the
calculation is to define the regularized charge of the ground state using the Atiyah-Patodi-Singer (APS) \emph{eta invariant} \cite{APS1} of the
\emph{spatial} Dirac operator $\mathcal{H}_{A,m}$. Note that in \cite{NS} the APS eta invariant was also referred to as the 
\emph{spectral asymmetry}, since the the APS eta invariant of a differential operator is a regularized version of the difference between the 
numbers of positive and negative eigenvalues of that operator. We also note that the calculation of the ground state charge in this 
subsection is quite general, and would also apply to the massive Dirac fermion on a general $D$-dimensional space. Thus, although our notation
is specialized to the case of $D=2$, the final result of Eq.~\eqref{eq:APS-charge} is also valid for spatial dimensions $D\neq 2$ as well. 

The operator $ \mathcal{H}_{A,m}$ has discrete\footnote{The eigenvalues are discrete because $\M$ is a closed 
manifold.} 
eigenvalues $E_n$ with corresponding eigenfunctions $\Phi_n(\mb{x})$, where $n$ is an index labeling the different eigenfunctions. 
The differential operator $\mathcal{H}_{A,m}$ is self-adjoint with respect to the inner product
\beq
	(\phi,\psi)= \int d^2 \mb{x} \ \sqrt{\text{det}[g(\mb{x})]} \phi^{\dg}(\mb{x})\psi(\mb{x})\ ,
\eeq
and we assume that the eigenfunctions $\Phi_n$ are orthonormal with respect to this inner product,
\beq
	(\Phi_n,\Phi_{n'})= \delta_{nn'}\ .
\eeq
Then the fermion operators can be defined by the mode expansion
\beq
	\hat{\Psi}(\mb{x})= \sum_{n} \hat{b}_n \Phi_n(\mb{x})\ , \label{eq:mode-expansion}
\eeq
where $\hat{b}_n$ are fermionic annihilation operators with the standard anticommutation relations 
$\{\hat{b}_n, \hat{b}^{\dg}_{n'}\}= \delta_{nn'}$. Using this mode expansion, we find that the Hamiltonian operator takes the diagonal form
\beq
	\hat{H}_{A,m}= \sum_{n} E_n \hat{b}^{\dg}_n \hat{b}_n\ .
\eeq

We now define the ground state $|0\ran_{A,m}$ for this system, corresponding to a Fermi (or Dirac) sea filled up to the energy $E=0$. In 
the case that $\mathcal{H}_{A,m}$ has zero modes, we have a choice about whether to keep 
those states empty or filled when we define the ground state $|0\ran_{A,m}$. 
For mathematical reasons that we discuss below, we 
choose to leave the zero energy states \emph{empty} in the state $|0\ran_{A,m}$. 
Note also that once we have computed the regularized charge of the ground
state $|0\ran_{A,m}$, the charges of all other states will be well-defined and will differ from the charge of $|0\ran_{A,m}$ by integer amounts.
This follows from the fact that we can obtain all of the other states by acting on $|0\ran_{A,m}$ with the $\hat{b}_n$ and $\hat{b}_n^{\dg}$ 
operators, which add or remove charge $1$ from the state $|0\ran_{A,m}$. 

With these considerations in mind, we now define the ground state $|0\ran_{A,m}$ by the conditions
\begin{subequations}
\beqa
	\hat{b}_n |0\ran_{A,m} &=& 0,\ E_n \geq 0\ , \\
	\hat{b}^{\dg}_n |0\ran_{A,m} &=& 0,\ E_n < 0\ ,
\eeqa
\end{subequations}
i.e., $|0\ran_{A,m}$ has all states with $E_n <0$ occupied. The charge in the ground state $|0\ran_{A,m}$ is then given by
\beq
	Q_{A,m}= {}_{A,m}\lan 0|\hat{Q}|0\ran_{A,m}\ ,
\eeq
where $\hat{Q}$ is the normal-ordered charge operator from Eq.~\eqref{eq:NN-charge}. If we plug in the mode expansion for
$\hat{\Psi}(\mb{x})$ into this expression for $Q_{A,m}$, then after some algebra we find the ill-defined expression
\beq
	Q_{A,m} = -\frac{1}{2} \left[ \left(\sum_{n;\ E_n >0} 1 - \sum_{n;\ E_n <0} 1\right) + h  \right]\ , 
\eeq
where
\beq
	h:= \text{dim}[\text{Ker}[ \mathcal{H}_{A,m}]]
\eeq
is the number of zero modes of $\mathcal{H}_{A,m}$. 

As discussed by Paranjape and Semenoff~\cite{PS} (and then used later by Niemi and Semenoff in \cite{NS}), 
it is possible to make sense of this expression by defining a regularized version of it using the 
APS \emph{eta invariant} of the \emph{spatial}\footnote{This is not the same as the eta invariant of the \emph{spacetime} 
Dirac operator that 
appears in path integral treatments of the parity anomaly~\cite{anomalies-odd,witten-fermions,seiberg-witten,witten-parity}.} 
Dirac operator $\mathcal{H}_{A,m}$. Recall that the \emph{eta function} $\eta(s)$ associated with $\mathcal{H}_{A,m}$ 
is~\cite{APS1}
\beq
	\eta(s)= \sum_{n; E_n\neq 0} \text{sgn}(E_n) |E_n|^{-s}\ ,
\eeq
which is an analytic function of $s\in\mathbb{C}$ when the real part of $s$ is sufficiently large. It is a nontrivial fact that $\eta(s)$ possesses
a well-defined analytic continuation to $s=0$. This analytic continuation is known as the APS eta invariant and it is denoted by $\eta(0)$. 
Following \cite{PS,NS}, we can now use $\eta(0)$ to define the regularized difference of the 
numbers of positive and negative eigenvalues of $\mathcal{H}_{A,m}$ as
\beq
	\left(\sum_{n;\ E_n >0} 1 - \sum_{n;\ E_n <0} 1\right)_{\text{reg.}}= \eta(0)\ .
\eeq

Using this regularization scheme, we find that the charge of the ground state $|0\ran_{A,m}$ is given by
\beq
	Q_{A,m}= -\frac{1}{2}\left(\eta(0) + h\right)\ . \label{eq:APS-charge}
\eeq
Note that if we had instead decided to define the ground state $|0\ran_{A,m}$ as having the zero modes all filled, then this would be
modified to 
\beq
	Q_{A,m}= -\frac{1}{2}\left(\eta(0) - h\right)\ .
\eeq
The mathematical reason for choosing the ground state $|0\ran_{A,m}$ to have all zero modes empty is that the combination 
\beq
	\eta(0)+h \nnb
\eeq
is exactly the combination that appears in the APS index theorem (Theorem 3.10 of Ref.~\onlinecite{APS1}). This means that if
if we choose to define $|0\ran_{A,m}$ in this way, then the APS index theorem can be applied to compute the ground state charge 
$Q_{A,m}$ in various systems that we might want to study. As we remarked above, once we have computed an appropriate 
regularized charge for the state $|0\ran_{A,m}$, the charges of all other states in the Hilbert space are well-defined and differ from
$Q_{A,m}$ by integer amounts.

\subsection{Ground state charge of the 2D Dirac fermion}

We now compute $\eta(0)+h$ for the Dirac fermion in 2D with Hamiltonian $\hat{H}_{A,m}$. This will give us the ground state charge 
$Q_{A,m}$ within the regularization scheme of \cite{NS}. To compute $\eta(0)+h$, first note that since 
$\{ \slashed{\nabla}_A, \overline{\gamma}\}=0$, we have
\beq
	\mathcal{H}_{A,m}^2= ( i\slashed{\nabla}_A)^2 + m^2\ ,
\eeq
which means that $\mathcal{H}_{A,m}$ has no zero modes, and so $h=0$. 
Next, we consider the calculation of $\eta(0)$ for $\mathcal{H}_{A,m}$. Let 
$\ep_n$ and $\phi_n(\mb{x})$ be the eigenvalues and eigenfunctions of the massless spatial Dirac operator $i\slashed{\nabla}_A$,
\beq
	 i\slashed{\nabla}_A \phi_n(\mb{x})= \ep_n \phi_n(\mb{x})\ .
\eeq
Then for eigenfunctions $\phi_n(\mb{x})$ with \emph{non-zero} eigenvalue $\ep_n$, we have
\begin{subequations}
\beqa
	\mathcal{H}_{A,m}\phi_n(\mb{x}) &=& \ep_n \phi_n(\mb{x}) + m \ov{\gamma}\phi_n(\mb{x}) \\
\mathcal{H}_{A,m}\ov{\gamma}\phi_n(\mb{x}) &=&  m \phi_n(\mb{x}) - \ep_n \ov{\gamma}\phi_n(\mb{x})\ .
\eeqa
\end{subequations}
By diagonalizing the $2\times 2$ matrix 
\beq
	\begin{pmatrix}
	\ep_n & m \\
	m & -\ep_n \\
	\end{pmatrix}\ ,
\eeq
we see that for any non-zero $\ep_n$, the massive Dirac Hamiltonian 
$\mathcal{H}_{A,m}$ has eigenvalues $\pm \sqrt{\ep_n^2 + m^2}$. These cancel each other in the computation of 
the eta invariant $\eta(0)$ of $\mathcal{H}_{A,m}$, so this means that only zero modes of  $i\slashed{\nabla}_A$ will 
contribute to $\eta(0)$. We consider these zero modes next.

As is well known, zero modes of $ i\slashed{\nabla}_A$ can be chosen to be eigenvectors of the chirality matrix
$\overline{\gamma}$ with eigenvalue (\emph{chirality}) equal to $\pm 1$. It is now easy to see that zero modes of 
$ i\slashed{\nabla}_A$ that have chirality $\pm 1$ are also eigenfunctions of $\mathcal{H}_{A,m}$ with eigenvalue $\pm m$. 
Let us assume for the moment that $m>0$. Then we find that the eta invariant for $\mathcal{H}_{A,m}$ reduces in this case to 
\beqa
	\eta(0) &=& (\#\ \text{of positive chirality zero modes of}\  i\slashed{\nabla}_A) \nnb \\
	&-& (\#\ \text{of negative chirality zero modes of}\  i\slashed{\nabla}_A)   \nnb \\
	&=& \text{Index}[ i\slashed{\nabla}_A]\ ,
\eeqa
where the last line follows from the \emph{definition} of $\text{Index}[ i\slashed{\nabla}_A]$, 
the index of the massless Dirac operator $i\slashed{\nabla}_A$.
An application of the Atiyah-Singer index theorem~\cite{AS} (a useful reference for physicists is Ref.~\onlinecite{EGH}) 
for the Dirac operator $i\slashed{\nabla}_A$ on the closed spatial manifold $\mathcal{M}$ then gives
\beq
	\text{Index}[ i\slashed{\nabla}_A]= \frac{1}{2\pi}\int_{\mathcal{M}} F\ , \label{eq:AS}
\eeq
where $F= \frac{1}{2}F_{ij}dx^i \wedge dx^j= dA$ is the field strength for the spatial gauge field $A=A_j dx^j$. 

It is important to note here that we have assumed that the background field $F$ obeys a \emph{Dirac quantization condition}, 
which states that the flux of $F$ through $\M$ must be an integer multiple of $2\pi$, 
\beq
	 \frac{1}{2\pi}\int_{\mathcal{M}} F \in \mathbb{Z}\ .
\eeq
Since the index of $i\slashed{\nabla}_A$ is integer-valued by definition, it is clear that Eq.~\eqref{eq:AS} would not make sense without 
this condition. Mathematically, this condition is equivalent to the statement that $A$ is a connection on a complex line bundle over 
$\M$, and the integer $ (2\pi)^{-1}\int_{\mathcal{M}} F$ is the first \emph{Chern number} of this line bundle (see, for example, Sec.~6 of
\cite{EGH}).

As a final comment on the Atiyah-Singer index theorem, we note that the sign on the right-hand side of Eq.~\eqref{eq:AS} can be seen to 
be correct by considering a simple example with $\mathcal{M}=S^2$,
the unit two-sphere. In this case we have $\int_{\M}d\omega=4\pi$ by the Gauss-Bonnet theorem (the Euler characteristic of $S^2$
is $2$). If we consider the field
configuration $A=\frac{1}{2}\omega$, then we
have $(2\pi)^{-1}\int_{\M}F=1$, and we can also see from Eq.~\eqref{eq:Dirac-A} that 
for this choice of $A$ the operator $i\slashed{\nabla}_A$ has a zero mode equal to a 
constant function on $\M$ times the eigenvector of $\ov{\gamma}$ with eigenvalue $+1$. This confirms that the sign in 
Eq.~\eqref{eq:AS} is correct. 

Using the result from the Atiyah-Singer index theorem \eqref{eq:AS}, we find that within this regularization scheme
the ground state charge for this system, for $m>0$, is
\beq
	Q_{A,m>0} = -\frac{1}{4\pi}\int_{\mathcal{M}} F= -\frac{1}{2}\frac{1}{2\pi}\int_{\M} F \ .
\eeq
This is in agreement with the result of Niemi and Semenoff who considered the case of 
flat space~\cite{NS}. We see that curving the space does not change the result. This is true because the 
Atiyah-Singer index theorem for the Dirac operator in 2D shows that the index of the operator $i\slashed{\nabla}_A$ 
does not receive any gravitational contribution (this is \emph{not} true in higher dimensions). The connection of the expression for 
$Q_{A,m}$ to the Atiyah-Singer index theorem was pointed out by Jackiw in Ref.~\onlinecite{jackiw1984}.

The above result was derived under the assumption that $m>0$. If we instead chose $m<0$, 
then our expression for the ground state charge would change sign because we would instead 
find that $\eta(0)= -\text{Index}[ i\slashed{\nabla}_A]$. Therefore, in the general case we find that
\beq
	Q_{A,m}= -\frac{\text{sgn}(m)}{2} \frac{1}{2\pi}\int_{\M} F\ . \label{eq:ground-q}
\eeq
An important property of this formula is that when $F$ is an odd multiple of $2\pi$, we find a half-integer charge 
in the ground state. This can occur, for example, if the Dirac fermion theory is located on the surface
of the TI (i.e., $\M$ is the surface of the TI) and if there is a magnetic monopole of the background electromagnetic field present in the bulk of 
the TI. In this case there would be a flux of $2\pi$ passing through $\M$, and our result for $Q_{A,m}$ shows that the ground state of the 
surface theory with the mass term $m\ov{\gamma}$ would have a charge of $\pm\frac{1}{2}$ depending on the sign of $m$.

\subsection{Discussion on symmetries}

We now explain that the regularization scheme of \cite{NS}, which we have been studying in this section, violates large $U(1)$ gauge invariance,
but is consistent with the time-reversal symmetry of the massless Dirac fermion. 
The violation of large $U(1)$ gauge invariance is easy to see from the
fact that the charge $Q_{A,m}$ from Eq.~\eqref{eq:ground-q} can take on \emph{half-integer} values. We now explain the sense in which
this regularization scheme is consistent with time-reversal symmetry. 

Recall that time-reversal acts on the Hamiltonian 
$\hat{H}_{A,m}$ as $\hat{T}\hat{H}_{A,m}\hat{T}^{-1}=\hat{H}_{-A,-m}$, 
i.e., the effect of time-reversal is to negate $A$ and $m$. Within the regularization scheme of \cite{NS}, which uses the eta
invariant of the spatial Dirac operator $\mathcal{H}_{A,m}$ to define $Q_{A,m}$, we find using Eq.~\eqref{eq:ground-q} that $Q_{A,m}$ 
satisfies the relation 
\beq
	Q_{A,m}= Q_{-A,-m}\ . \label{eq:GS-charge-TR-section}
\eeq
This means that in this regularization scheme the ground state charge for the theory with Hamiltonian $\hat{H}_{A,m}$ is equal to the
ground state charge of the \emph{time-reversed} theory with Hamiltonian $\hat{H}_{-A,-m}$. This is the precise sense
in which the eta invariant regularization scheme of \cite{NS} is consistent with the time-reversal symmetry of the massless Dirac fermion.  

One way to understand why Eq.~\eqref{eq:GS-charge-TR-section} holds within this regularization is to note that the eta 
invariant is built from the spectrum of the massive Dirac operator $\mathcal{H}_{A,m}$, and 
$\mathcal{H}_{A,m}$ and $\mathcal{H}_{-A,-m}$ have the
\emph{same} spectrum. To see this, observe that if $\Phi(\mb{x})$ is an eigenfunction of $\mathcal{H}_{A,m}$ with eigenvalue
$E$, then $\ov{\gamma}\Phi^*(\mb{x})$ is an eigenfunction of $\mathcal{H}_{-A,-m}$ with
the same eigenvalue (the star $^*$ denotes complex conjugation). 

\section{Regularization scheme 2}
\label{sec:reg2}

In this section we study the Dirac fermion using our second regularization scheme, which is a \emph{lattice} regularization scheme for the
Dirac fermion on a spatial torus. In this regularization scheme we compute the Hall conductivity $\sigma_{H,m}$ in the ground state of the 
Dirac fermion with mass $m$ and in the presence of a background time-independent electric field $\mb{E}$. We find that $\sigma_{H,m}$ 
is always an integer (in units of $\frac{e^2}{h}$), which implies that this regularization scheme preserves the large $U(1)$ gauge invariance of the 
Dirac fermion (there are no fractionalized quantum numbers found in the Hall response of the system to the background electric field). 
On the other hand, we show that this regularization scheme explicitly breaks the 
time-reversal symmetry of the massless Dirac fermion in the continuum. This fact is also reflected in the result of the Hall conductivity calculation, 
where we find that $\sigma_{H,m}\neq -\sigma_{H,-m}$. 

As we mentioned in the Introduction, the calculation in this section is closely related to the calculation of 
Haldane~\cite{haldane-parity} on a lattice model on the honeycomb lattice that displays a non-zero Hall conductivity in the absence of any net 
external magnetic field (i.e., zero total magnetic flux through each unit cell). 
Our results here are consistent with the findings in Ref.~\onlinecite{haldane-parity}. Just as in Haldane's
model on the honeycomb lattice, the model that we consider also features a single massless Dirac fermion at low 
energies, but at the cost of breaking time-reversal symmetry. In fact, it was emphasized in 
Ref.~\onlinecite{haldane-parity} that the honeycomb model considered there 
should be thought of as a condensed matter realization of the parity anomaly.

\subsection{Lattice regularization and Hall conductivity}

For the lattice regularization we consider a set of two-component fermions on the square lattice and with periodic boundary conditions, and
we set the lattice spacing equal to $1$. The Fourier transform of the two-component lattice fermion operator will be denoted by
$\hat{\Psi}(\mb{k})$, with components $\hat{\Psi}_{\al}(\mb{k})$, $\al=1,2$. Here $\mb{k}=(k_1,k_2)$ is a wave vector in the
first Brillouin zone of the square lattice, $\mb{k}\in (-\pi,\pi]\times(-\pi,\pi]$. The Hermitian conjugate of $\hat{\Psi}(\mb{k})$
is $\hat{\Psi}^{\dg}(\mb{k})$ with components $\hat{\Psi}^{\dg,\al}(\mb{k})$, $\al=1,2$. 
We take the Hamiltonian for the lattice regularization of the Dirac fermion to be
\beq
	\hat{H}_{\text{lattice}}= \sum_{\mb{k}}\hat{\Psi}^{\dg}(\mb{k})\mathcal{H}(\mb{k})\hat{\Psi}(\mb{k})\ ,
\eeq
where the Bloch Hamiltonian $\mathcal{H}(\mb{k})$ is given by
\beq
	\mathcal{H}(\mb{k})= \sin(k_1)\sigma^x + \sin(k_2)\sigma^z + (\td{m}+2-\cos(k_1)-\cos(k_2))\sigma^y\ , \label{eq:Bloch-Ham}
\eeq 
Here $\td{m}$ is a tunable parameter that, in a certain parameter regime,
can be identified with the mass $m$ of the continuum Dirac fermion. This model features two bands, labeled ``$+$'' and ``$-$'', with
energies given by $\mathcal{E}_{\pm}(\mb{k})= \pm \lambda(\mb{k})$ with
\beq
	\lambda(\mb{k})= \sqrt{ \sin^2(k_1) + \sin^2(k_2) + (\td{m}+2-\cos(k_1)-\cos(k_2))^2 }\ . \label{eq:def-lambda}
\eeq
In what follows we will be interested in the case in which the lower band is completely filled and the upper band is completely empty. 
We also note here that essentially the same model was studied in 
Sec.~II.B of Ref.~\onlinecite{QHZ}.

Consider the parameter regime $|\td{m}| \ll 1$. In this regime the upper and lower bands of the model come closest to each other at the
origin $\mb{k}=(0,0)$ of the Brillouin zone, and the two bands actually touch at $\mb{k}=(0,0)$ when $\td{m}=0$. 
If we Taylor expand the Bloch Hamiltonian $\mathcal{H}(\mb{k})$ near $\mb{k}=(0,0)$, then we find that it takes the approximate 
form
\beq
	\mathcal{H}(\mb{k})\approx k_1\sigma^x + k_2\sigma^z + \td{m} \sigma^y\ . \label{eq:FT-Dirac}
\eeq
To make contact with our previous discussion of the Dirac operator in the continuum, recall that we worked in a basis in 
which the gamma matrices $\gamma^a$, $a=1,2$, were both real, and so the third matrix $\ov{\gamma}$ was imaginary.
One concrete choice for these matrices is $\gamma^{1}= \sigma^x$, $\gamma^{2}=\sigma^z$, which gives
$\ov{\gamma}= \sigma^y$. With this choice, we see that the Fourier transform of the massive Dirac operator
$i\slashed{\nabla}+m\ov{\gamma}$ on flat space has exactly the form of Eq.~\eqref{eq:FT-Dirac} with
\beq
	m=\td{m}\ .
\eeq

The discussion in the previous paragraph shows that in the regime $|\td{m}|\ll 1$, the low energy description of this lattice model 
consists of a single continuum Dirac fermion with mass $m=\td{m}$ and located at the point $\mb{k}=(0,0)$ in the Brillouin zone of the square 
lattice. In addition, the full lattice model does not
have any additional phase transitions\footnote{By a phase transition we mean a value of the parameter $\td{m}$ 
at which the upper and lower bands touch.} for any $\td{m}>0$, while the next transition
for $\td{m}<0$ occurs at $\td{m}=-2$. At $\td{m}=-2$ the upper and lower bands touch at the two points $\mb{k}=(\pi,0)$ and 
$\mb{k}=(0,\pi)$. This means that this lattice model is a sensible regularization for a single continuum Dirac fermion as long as we keep 
the parameter $\td{m}$ in a region near $\td{m}=0$ and far away from the next transition at $\td{m}=-2$. 

We now turn to the calculation of the Hall conductivity for the Dirac fermion in this lattice regularization. We first compute the Hall
conductivity $\sigma_{H,\td{m}}^{\text{{\tiny{lattice}}}}$ for the lattice model, which is well-defined for any value of the parameter 
$\td{m}$
for which there is a gap between the upper and lower bands of the model. We then identify the Hall conductivity 
$\sigma_{H,m}$ of the continuum Dirac fermion with the lattice Hall conductivity $\sigma_{H,\td{m}}^{\text{{\tiny{lattice}}}}$ in the appropriate
parameter regime where the lattice model is a sensible regularization of the continuum Dirac fermion. Specifically, we have the following
identifications,
\begin{subequations}
\label{eq:lattice-continuum-hall}
\beqa
	\sigma_{H,m>0} &=& \sigma_{H,\td{m}>0}^{\text{{\tiny{lattice}}}} \\
	\sigma_{H,m<0} &=& \sigma_{H,-2<\td{m}<0}^{\text{{\tiny{lattice}}}}\ .
\eeqa
\end{subequations}

The Hall conductivity $\sigma_{H,\td{m}}^{\text{{\tiny{lattice}}}}$ for the lattice model is defined precisely as follows. We 
first place the
system in a static electric field $\mb{E}$ that points in the $x^2$ direction, so that $\mb{E}=(0,E_2)$. We then compute the
current $j^1$ that flows in the $x^1$ direction. Then $\sigma_{H,\td{m}}^{\text{{\tiny{lattice}}}}$ is defined as the constant
that relates $j^1$ to $E_2$, 
\beq
	j^1 = \frac{\sigma_{H,\td{m}}^{\text{{\tiny{lattice}}}}}{2\pi}E_2\ .
\eeq
More precisely, $\sigma_{H,\td{m}}^{\text{{\tiny{lattice}}}}$ encodes the spatially uniform (i.e., zero wave vector) part of the
\emph{linear response} of $j^1$ to the applied field $E_2$.
Note also that the factor of $(2\pi)^{-1}$ appearing here is actually $\frac{e^2}{h}$ in our units where $e=\hbar=1$.

As discussed above, we consider the case
where the lower band is completely filled and the upper band is completely empty, as this filling corresponds to the continuum ground state in 
which the Dirac sea of negative energy states is completely filled. 
In this case we can compute $\sigma_{H,\td{m}}^{\text{{\tiny{lattice}}}}$ using various methods 
including a direct linear response calculation using the Kubo formula~\cite{thouless1982quantized}, or the semi-classical theory of wave packet 
dynamics in solids~\cite{niu-review}. Both methods lead to the result that
\beq
	\sigma_{H,\td{m}}^{\text{{\tiny{lattice}}}}= -\int \frac{d^2\mathbf{k}}{2\pi}\Omega^{12}_{-}(\mathbf{k})\ ,
\eeq
where $\Omega^{j\ell}_{-}(\mathbf{k})$ (with $j,\ell=1,2$) 
are the components of the \emph{Berry curvature} of the filled lower band (the ``$-$'' band) 
of the lattice model, and where the integral is taken over the Brillouin zone of the square lattice. 
The Berry curvatures $\Omega^{j\ell}_{\pm}(\mathbf{k})$ for the
``$+$'' and ``$-$'' bands of the model are defined precisely as follows.
Let $|u_{\mb{k},\pm}\ran$ be the eigenvector of the Bloch Hamiltonian corresponding to the $\pm$ band of the model, 
\beq
	\mathcal{H}(\mb{k})|u_{\mb{k},\pm}\ran= \mathcal{E}_{\pm}(\mb{k})|u_{\mb{k},\pm}\ran\ .
\eeq
If we define the \emph{Berry connection} for the $\pm$ band as 
$\mathcal{A}^j_{\pm}(\mb{k})= i\Big\lan u_{\mb{k},\pm}\Big| \frac{\pd u_{\mb{k},\pm}}{\pd k_j}\Big\ran$, then the 
Berry curvature for the $\pm$ band is given by
\beq
	\Omega^{j\ell}_{\pm}(\mb{k}) = \frac{\pd\mathcal{A}^{\ell}_{\pm}(\mb{k})}{\pd k_j}-\frac{\pd\mathcal{A}^j_{\pm}(\mb{k})}{\pd k_{\ell}}\ .
\eeq

We now provide some details of the Berry curvature calculation. For this calculation it is convenient 
to introduce spherical coordinate variables $\Theta(\mb{k})$ and $\Phi(\mb{k})$ and to rewrite the 
Bloch Hamiltonian in terms of these variables as
\begin{align}
	\mathcal{H}(\mb{k})= \lambda\Big( \sin(\Theta)\cos(\Phi)\sigma^x + \sin(\Theta)\sin(\Phi)\sigma^y + \cos(\Theta)\sigma^z \Big)\ ,
\end{align}
where $\lambda(\mb{k})$ was defined in Eq.~\eqref{eq:def-lambda}, and where we have suppressed the dependence of 
$\lam(\mb{k}),\Theta(\mb{k})$, and $\Phi(\mb{k})$ on $\mb{k}$
for brevity. This type of parametrization for a two-band Hamiltonian has been used, for example, in 
Sec.~I.C.3 of Ref.~\onlinecite{niu-review}. In terms of these variables the eigenvector $|u_{\mb{k},-}\ran$ for the 
lower band of the model takes the form
\beq
	|u_{\mb{k},-}\ran= \begin{pmatrix}
	e^{-i\Phi}\sin\left(\frac{\Theta}{2}\right) \\
	-\cos\left(\frac{\Theta}{2}\right)
	\end{pmatrix}\ .
\eeq
A straightforward calculation then shows that the Berry curvature for the lower band is given by
\beq
	\Omega^{12}_{-}(\mb{k})= -\frac{1}{2}\ep_{j\ell}\frac{\pd \Phi(\mb{k})}{\pd k_j}\frac{\pd \Theta(\mb{k})}{\pd k_{\ell}}\sin(\Theta(\mb{k}))\ ,
\eeq
and so we have
\beq
	\sigma_{H,\td{m}}^{\text{{\tiny{lattice}}}}= \frac{1}{4\pi}\int d^2\mb{k}\ \ep_{j\ell}\frac{\pd \Phi(\mb{k})}{\pd k_j}\frac{\pd \Theta(\mb{k})}{\pd k_{\ell}}\sin(\Theta(\mb{k}))\ . \label{eq:Hall-integral}
\eeq
This expression shows that $\sigma_{H,\td{m}}^{\text{{\tiny{lattice}}}}$ is an integer and is equal to the number
of times that the unit vector specified by $\Theta(\mb{k})$ and $\Phi(\mb{k})$ covers the unit two-sphere $S^2$ as 
$\mb{k}$ varies over the Brillouin zone of the square lattice. This can be seen from the fact that 
$\sin(\Theta) d\Theta d\Phi$ is the area element on $S^2$, and from the fact that
$\ep_{j\ell}\frac{\pd \Phi(\mb{k})}{\pd k_j}\frac{\pd \Theta(\mb{k})}{\pd k_{\ell}}$ is the Jacobian of the map from the Brillouin zone to 
$S^2$ (the normalizing factor of $4\pi$ is also the total area of $S^2$).

One way to proceed with the calculation of $\sigma_{H,\td{m}}^{\text{{\tiny{lattice}}}}$ would be to work out
explicit expressions for $\Theta(\mb{k})$ and $\Phi(\mb{k})$ in terms of $\mb{k}$ and $\td{m}$ and then evaluate
the integral in Eq.~\eqref{eq:Hall-integral}. As a practical matter, however, the easiest way to compute 
$\sigma_{H,\td{m}}^{\text{{\tiny{lattice}}}}$ is to evaluate the integral numerically for a particular value of 
$\td{m}$ in each parameter range where the Hamiltonian has a gap between the upper and lower bands. We can 
use this method because we already know that $\sigma_{H,\td{m}}^{\text{{\tiny{lattice}}}}$ is an integer-valued topological invariant
that takes a constant value in each parameter range where the Hamiltonian has a gap. 
We find that the Hall conductivity for the lattice model is given by
\beq
	\sigma_{H,\td{m}}^{\text{{\tiny{lattice}}}}=\begin{cases}
		0 \ , & \ \td{m}>0 \\
		-1 \ , & -2 < \td{m} < 0 \\
		 1\ , & -4 < \td{m} <-2 \\
		 0 \ , & \td{m} < -4
	\end{cases}\ .
\eeq
Only the first two cases $\td{m}>0$ and $-2<\td{m}<0$ are relevant for our original goal of studying a regularization of the continuum Dirac 
fermion. Using Eq.~\eqref{eq:lattice-continuum-hall} to compute $\sigma_{H,m}$ from $\sigma_{H,\td{m}}^{\text{{\tiny{lattice}}}}$, we find 
that the result for $\sigma_{H,m}$, for either sign of the mass $m$, can be written in the compact form
\beq
	\sigma_{H,m}= \frac{\text{sgn}(m)-1}{2}\ . \label{eq:Hall-m}
\eeq

\subsection{Discussion on symmetries}

We now point out that this lattice regularization preserves the large $U(1)$ gauge invariance of the original massless Dirac fermion, but 
breaks the time-reversal symmetry. To see that large $U(1)$ gauge invariance is preserved, it is sufficient to note that this regularization yields an 
integer value for the Hall conductivity $\sigma_{H,m}$. In other words, we do not find any fractionalization of quantum numbers associated with 
the $U(1)$ symmetry of charge conservation. This result makes sense since in this regularization scheme we are dealing with a well-defined lattice 
model with charge conservation symmetry. 

We now show that the lattice regularization that we have been discussing does not possess the time-reversal symmetry of the
continuum massless Dirac fermion, even when the mass parameter $\td{m}$ in the lattice model is set to zero. To see this, note that 
for the choice $\gamma^{1}= \sigma^x$, $\gamma^{2}=\sigma^z$, and $\ov{\gamma}= \sigma^y$, the time-reversal operator defined
in Eq.~\eqref{eq:TR-def} would act on the lattice fermions $\hat{\Psi}(\mb{k})$ as 
\begin{subequations}
\label{eq:TR-def-lattice}
\beqa
	\hat{T} \hat{\Psi}_{\al}(\mb{k}) \hat{T}^{-1} &=& {(\sigma^y)_{\al}}^{\beta}\hat{\Psi}_{,\beta}(-\mb{k}) \\
	\hat{T} \hat{\Psi}^{\dg,\al}(\mb{k}) \hat{T}^{-1} &=& \hat{\Psi}^{\dg,\beta}(-\mb{k}) {(\sigma^y)_{\beta}}^{\al}\ ,
\eeqa 
\end{subequations}
where we note that $\mb{k}$ is negated by the time-reversal operation. 
Then the condition of time-reversal invariance of the Hamiltonian, $\hat{T}\hat{H}_{\text{lattice}}\hat{T}^{-1}=\hat{H}_{\text{lattice}}$,
is equivalent to the matrix equation
\beq
	\sigma^y\mathcal{H}^*(\mb{k})\sigma^y= \mathcal{H}(-\mb{k})\ .
\eeq
However, it is easy to check that this condition is \emph{not} satisfied by $\mathcal{H}(\mb{k})$, even when $\td{m}=0$. 
It follows that this lattice regularization explicitly breaks the time-reversal symmetry of the continuum massless Dirac fermion. 

Finally, we note that the breaking of time-reversal symmetry in this regularization scheme
can also be seen from the fact that 
\beq
	\sigma_{H,m}\neq -\sigma_{H,-m}\ ,
\eeq
i.e., time-reversed theories \emph{do not} have opposite values of the Hall conductivity within this regularization 
scheme.\footnote{The electric field $\mb{E}$ is invariant under time-reversal, so the time-reversal partner of the theory with mass $m$
in the presence of $\mb{E}$ is the theory with mass $-m$ in the presence of the same field $\mb{E}$.}

\section{Effective actions for the two regularizations}
\label{sec:effective-action}

In this final section we compute, for each regularization scheme, the effective action $S_{\text{eff}}[A]$ that encodes the response of the system
to the background electromagnetic field $A=A_{\mu}dx^{\mu}$. On curved space the physical three-current $j^{\mu}(x)$ 
that arises as a response to the background field $A$ is obtained from $S_{\text{eff}}[A]$ by functional differentiation as
\beq
	j^{\mu}(x)= -\frac{1}{\sqrt{\text{det}[g(\mb{x})]}}\frac{\delta S_{\text{eff}}[A]}{A_{\mu}(x)}\ ,
\eeq
where we remind the reader that $x=(x^0,x^1,x^2)$ is the spacetime coordinate and $\mb{x}=(x^1,x^2)$ is the spatial coordinate. The
overall minus sign appearing here is a matter of convention. We chose it because with this sign the $j^0(x)$ term in the effective action
has the form
\beq
	- \int d^3x\ \sqrt{\text{det}[g(\mb{x})]} j^0(x)A_0(x)\ , \nnb
\eeq 
which has the correct sign for the action arising from the potential energy of the charge density $j^0$ in the presence of the 
scalar electromagnetic potential $A_0$.

For both regularizations considered in this note, we find that the effective action that encodes the response has the Chern-Simons form
\beq
	S_{CS}[A]= \frac{k}{4\pi}\int A\wedge dA= \frac{k}{4\pi}\int d^3x\ \ep^{\mu\nu\lam}A_{\mu}\pd_{\nu}A_{\lam}\ ,
\eeq 
for an appropriate choice of the level $k$. To find the correct value of $k$ in each case, we compute the response that follows from 
$S_{CS}[A]$ in the two situations considered in this note. 

We start by computing the charge that follows from $S_{CS}[A]$ for a system on a closed spatial manifold $\M$ and in the presence of 
a time-independent spatial gauge field. This is exactly the physical quantity that we computed using the regularization scheme of 
Niemi and Semenoff in Sec.~\ref{sec:reg1}. This charge is given by
\beqa
	Q &=& \int d^2\mb{x}\ \sqrt{\text{det}[g(\mb{x})]} j^0(x) \nnb \\\
		&=& -\frac{k}{2\pi}\int_{\M}F\ ,
\eeqa
where in the second line we plugged in the result for $j^0(x)$ that follows from functional differentiation of $S_{CS}[A]$. 
To match with our answer for $Q_{A,m}$ from Eq.~\eqref{eq:ground-q}, we find that we must choose the level $k$ to be
\beq
	k_{\text{NS}}= \frac{\text{sgn}(m)}{2}\ .
\eeq
Then for the regularization scheme of Niemi and Semenoff we find the effective action
\beq
	S^{(\text{NS})}_{\text{eff}}[A]= \frac{\text{sgn}(m)}{2}\frac{1}{4\pi}\int A\wedge dA\ .
\eeq

Next, we compute the Hall conductivity for the case where space is a flat torus, which is exactly the physical quantity that we computed using
the lattice regularization scheme in Sec.~\ref{sec:reg2}. For this calculation we study the current that flows in the $x^1$ direction in response to a 
static electric field $\mb{E}=(0,E_2)$ pointing in the $x^2$ direction. For the effective action of the Chern-Simons form we find that
\beq
	j^1=\frac{k}{2\pi}E_2\ ,
\eeq
where we needed to use the equation $F_{20}= \pd_2 A_0 - \pd_0 A_2= - E_2$, which relates the physical electric field to the components of 
$A$ (see the end of Appendix~\ref{app:Dirac} for a review of this relation). This equation implies a Hall conductivity of $\sigma_H= k$. 
In this case, to match our answer for $\sigma_{H,m}$ from Eq.~\eqref{eq:Hall-m}, we must choose the level $k$ as
\beq
	k_{\text{lattice}}= \frac{\text{sgn}(m)-1}{2}\ .
\eeq
Then for the lattice regularization scheme on a spatial torus we find the effective action
\beq
	S^{(\text{lattice})}_{\text{eff}}[A]= \left(\frac{\text{sgn}(m)-1}{2}\right)\frac{1}{4\pi}\int A\wedge dA\ .
\eeq

It is now clear that the effective actions computed using the two different regularization schemes differ by a Chern-Simons counterterm as
\beq
	S^{\text{(lattice)}}_{\text{eff}}[A]= S^{\text{(NS)}}_{\text{eff}}[A] -\frac{1}{2}\frac{1}{4\pi}\int A\wedge dA\ ,
\eeq
where we see that the Chern-Simons counterterm has a \emph{fractionally-quantized} level equal to $-\frac{1}{2}$. This difference between the
effective actions for these two regularization schemes exactly matches the expectation from the original path integral treatment of the
parity anomaly~\cite{redlich1,redlich2}. 

\section{Conclusion}
\label{sec:con}

In this note we reviewed the parity anomaly of the massless Dirac fermion in $2+1$ dimensions in the context of the Hamiltonian
formalism, as opposed to the more conventional discussion within the path integral formalism. 
Our first goal with this presentation was to explain the parity anomaly in a way that would be more approachable for
condensed matter physicists. To this end, we have tried to show how the anomaly is manifested in the calculation of concrete 
physical quantities such as the charge of the ground state in a background spatial gauge field $A$ (Sec.~\ref{sec:reg1}) and the Hall conductivity 
of the ground state in a background electric field $\mb{E}$ (Sec.~\ref{sec:reg2}). 

Our second goal was to understand the precise relation between time-reversal symmetry and the charge of $\pm\frac{1}{2}$ that appears
on the surface of the TI (whose surface theory is the massless Dirac fermion) when a magnetic monopole is present in the bulk of the TI. 
The regularization scheme that leads to this half-quantized charge is known and was originally considered by Niemi and Semenoff in 
Ref.~\onlinecite{NS}. In this note we explained that this regularization scheme is consistent with the time-reversal symmetry of the 
massless Dirac fermion, in the precise sense of Eq.~\eqref{eq:GS-charge-TR}. 
To the best of our knowledge, the consistency of the regularization scheme of \cite{NS} with 
time-reversal symmetry has not been discussed in detail in the existing literature (see, however, the comparison with a 
parity-preserving point-splitting regularization scheme in Ref.~\onlinecite{barci2001point}).
This observation is important because it fits in with the
general picture of the parity anomaly, which states that a given regularization scheme can preserve either the time-reversal symmetry, or the
large $U(1)$ gauge invariance of the massless Dirac fermion, but not both. 

An interesting direction for future work on this topic would be to investigate a bosonic analogue of the parity anomaly in
quantum field theories with $U(1)$ and time-reversal symmetry that can appear on the surface 
of the bosonic topological insulator~\cite{VS2013,MKF1}. The bosonic topological insulator is the closest analogue, in a bosonic system with
$U(1)$ and time-reversal symmetry, of the more familiar fermion TI state. 
Some ideas about the form of this bosonic anomaly have already been presented in Ref.~\onlinecite{lapa1}. The
key physical property of the anomaly that was discussed there was the fact that a bosonic theory possessing this anomaly can be driven
into a time-reversal breaking state with a Hall conductivity of $1$ (in units of $\frac{e^2}{h}$), which is exactly \emph{half} of the 
allowed Hall conductivity that can be achieved in a (non-fractionalized) phase of bosons that can exist intrinsically in $2+1$ 
dimensions~\cite{lu-vishwanath,levin-senthil}. 

An additional reason to look for such an anomaly in $2+1$ dimensions is the demonstration in Ref.~\onlinecite{lapa2} that in $0+1$ dimensions 
there is a bosonic anomaly that is an exact analogue of a well-known fermionic anomaly in the same dimension~\cite{elitzur1986origins}.
The action for a massless Dirac fermion in $0+1$ dimensions coupled to a background gauge field $A=A_0dx^0$ has both
large $U(1)$ gauge invariance and a unitary \emph{charge conjugation} (or particle-hole) symmetry. However, it was shown in 
Ref.~\onlinecite{elitzur1986origins} that it is impossible to regularize this theory in a way that preserves both of these symmetries. 
This anomaly is clearly analogous to the parity anomaly of the massless Dirac fermion in $2+1$ dimensions, but with charge conjugation
instead of time-reversal as the relevant discrete symmetry. It was recently shown in Ref.~\onlinecite{lapa2} that
an exact analogue of this anomaly exists in a bosonic theory in $0+1$ dimensions with the \emph{same symmetries}. 
In addition, the calculation of the bosonic anomaly in 
Ref.~\onlinecite{lapa2} employed the \emph{equivariant localization} technique, which revealed that the bosonic and fermionic anomalies
in $0+1$ dimensions have the same mathematical origin. Indeed, the derivation showed that both anomalies follow from the form of the 
APS eta invariant for the Dirac operator in $0+1$ dimensions. Based on this simple example, we expect that it would be interesting
to search for a bosonic analogue of the parity anomaly in $2+1$ dimensions.

\acknowledgements

We thank M. Levin for helpful discussions on this topic and for a collaboration on a related project. 
M.F.L. acknowledges the support of the Kadanoff Center for Theoretical Physics at the University of Chicago.

\appendix

\section{Conventions}
\label{app:Dirac}

In this appendix we review our conventions and notation for the Dirac operator on curved space and our conventions for the
electromagnetic field. The information contained in this appendix is used in Sec.~\ref{sec:Dirac-fermion} of the main text of this note, where we
review the Hamiltonian for the Dirac fermion on flat and curved space. 

We consider a Dirac fermion on a spacetime of the form $\M\times\mathbb{R}$, where $\M$ represents $D$-dimensional space and 
$\mathbb{R}$ represents time. We assume that $\M$ is an orientable Riemannian manifold. We also need to assume that
$\M$ is a \emph{spin manifold} so that we can consistently place fermions on $\M\times\mathbb{R}$. We also assume that
$\M$ is closed (i.e., compact and without boundary) and connected.
Coordinates on the full spacetime will be denoted by $x^{\mu}$ where the (Greek) spacetime indices 
$\mu,\nu,\dots$ take the values $\{0,1,\dots,D\}$. The spatial coordinates on $\M$ are $x^j$ where the (Latin) spatial indices $j,k,\dots$ take the 
values $\{1,2,\dots,D\}$. We denote by $x=(x^0,\dots,x^D)$ the full vector of spacetime coordinates and by $\mb{x}=(x^1,\dots,x^D)$ the 
vector of spatial coordinates. We also use the standard summation convention in which we sum over any index (Latin or Greek) that is repeated 
once as a subscript and once as a superscript in any expression.

We denote by $G_{\mu\nu}$ the components of the spacetime metric 
$G$, which we choose to have signature $(1,-1,\dots,-1)$ (i.e., a ``mostly minus" signature). 
Since our spacetime is a product of a curved space $\M$ and flat time direction $\mathbb{R}$, the spacetime metric $G$ has the form
\beq
	G= dx^0\otimes dx^0 - g_{jk}(\mb{x})dx^j\otimes dx^k\ ,
\eeq
where $g_{jk}(\mb{x})$ are the components of an ordinary Riemannian metric $g$ on $\M$. Note that with this definition 
we have $\text{det}[g(\mb{x})]>0$ for all $\mb{x}\in\M$, where $\text{det}[g(\mb{x})]$ is the determinant of $g_{jk}(\mb{x})$.

We now discuss the construction of the spatial Dirac operator on $\M$. The first step is to define the
the coframe one-forms $e^a= e^a_{j}dx^j$ and frame vector fields $e_a= e^j_a \pd_j$ ($\pd_j\equiv \frac{\pd}{\pd x^j}$), where
frame indices $a,b,\dots$ take the values $\{1,\dots,D\}$. The components of 
these objects are defined in terms of the metric $g_{jk}$ by
\begin{subequations}
\beqa
	e^a_j\delta_{ab}e^b_k &=& g_{jk} \\
	e^{j}_a g_{jk} e^{k}_b &=& \delta_{ab}\ .
\eeqa
\end{subequations}
The frame and coframe components 
are inverses of each other when considered as matrices, $e^a_{j} e^{j}_b= \delta^a_b$ and 
$e^a_{k}e^{j}_a= \delta^{j}_{k}$. In addition, we have the relation 
$\text{det}[e(\mb{x})]= \sqrt{\text{det}[g(\mb{x})]}$, where $\text{det}[e(\mb{x})]$ is the determinant of the \emph{coframe} 
$e^a_{j}$ viewed as a matrix with row index $a$ and column index $j$. 

To construct a Dirac operator on $\M$ we also need a set of gamma matrices $\gamma^a$ with frame indices $a\in\{1,\dots,D\}$.
These satisfy the Clifford algebra $\{\gamma^a,\gamma^b\}= 2\delta^{ab}$, and in terms of them we define the rotation generators
$\gamma^{ab} := \frac{1}{2}[\gamma^a,\gamma^b]$ (these are generators of the group $\text{Spin}(D)$). Note that these
gamma matrices all square to the identity and so we can choose them to be Hermitian.

The next ingredient we need is the spin connection on $\M$. 
The spin connection one-form ${\omega^a}_b= {{\omega_{j}}^a}_b dx^j$ on $\M$ is defined by the relation
\beq
	\nabla_{j} e_b= {{\omega_{j}}^a}_b e_a\ ,
\eeq
where $\nabla_{j}\equiv \nabla_{\pd_j}$ denotes the connection on the tangent bundle of $\M$. Under a local rotation of the coframes
$e^a \to {\Lambda^a}_b e^b$, the spin connection transforms as
\beq
	{{\omega_{j}}^a}_b \to {(\Lambda \omega_{j} \Lambda^{-1})^a}_b - {(\pd_{j}\Lambda \Lambda^{-1})^a}_b\ .
\eeq
If we assume metric compatibility of the spin connection, then we have $\omega_{ab}= -\omega_{ba}$, i.e., $\omega_{ab}$ is a one-form that takes values in the Lie algebra of the group $SO(D)$.

The curvature and torsion two-forms on $\M$ are 
\begin{subequations}
\beqa
	{R^a}_b &=& d{\omega^a}_b + {\omega^a}_c \wedge {\omega^c}_b \\
	T^a &=& d e^a +  {\omega^a}_b \wedge e^b\ . 
\eeqa
\end{subequations}
If we assume that the torsion vanishes, ${T^a}_{jk}= 0$ for all $a,j,k$, where ${T^a}_{jk}$ are the components of the two-form 
$T^a= \frac{1}{2}{T^a}_{jk}dx^j\wedge dx^k$, 
then the metric-compatible spin connection takes the explicit form
\beq
	{{\omega_{j}}^a}_b= e^{k}_b\Gamma^{\ell}_{jk}e^a_{\ell} - e^{k}_b\pd_{j}e^a_{k} \ ,\label{eq:spin-connection}
\eeq
where $\Gamma^{\ell}_{jk}$ is the Levi-Civita connection,
\beq
	\Gamma^{\ell}_{jk}= \frac{1}{2}g^{\ell m}\left(\pd_{k}g_{m j} + \pd_{j}g_{m k} - \pd_{m} g_{jk} \right)\ .
\eeq

With all of these conventions in place, we can now construct the Dirac operator on the $D$-dimensional space $\M$. The
Dirac operator $\mathcal{D}$ is given by
\beq
	\mathcal{D}= i\slashed{\nabla}\ ,
\eeq
where
\beq
	\slashed{\nabla}= e^{j}_a\gamma^a(\pd_{j} + \frac{1}{4}\omega_{j b c}\gamma^{bc})\ . \label{eq:Dirac-op}
\eeq
One can check that $\mathcal{D}$ is Hermitian with respect to the inner product
\beqa
	(\psi,\phi) &=& \int d^D \mb{x}\ \text{det}[e(\mb{x})] \psi^{\dg}(\mb{x})\phi(\mb{x}) \nnb \\
		&=& \int d^D \mb{x}\ \sqrt{\text{det}[g(\mb{x})]} \psi^{\dg}(\mb{x})\phi(\mb{x})  \ ,
\eeqa
which is the appropriate inner product for spinors $\phi$ and $\psi$ on the Riemannian manifold $\M$. To verify that $\mathcal{D}$ is Hermitian 
one needs to use the fact that the torsion two-form vanishes, ${T^a}_{jk}= 0$ for all $a,j,k$, as well as the fact that $\M$ is closed 
(no boundary terms arise in integration by parts because $\M$ does not have a boundary).

To close this appendix we also discuss our conventions for the electromagnetic field. 
We specialize the discussion to the case of $D=2$, which is the case that we consider in the main text of this note.
Let $A=A_{\mu}dx^{\mu}$ be the one-form for a configuration of a background electromagnetic field. 
Since we assume a spacetime metric with signature $(1,-1,-1)$, when the space $\M$ is flat (i.e., $\M=\mathbb{R}^2$) the components 
$F_{\mu\nu}$ of 
the field strength two-form $F=dA= \frac{1}{2}F_{\mu\nu}dx^{\mu}\wedge dx^{\nu}$ are related to the usual electric and magnetic fields  as
\begin{subequations}
\beqa
	F_{12} &=& -B \\
	F_{20} &=& -E_2 \\
	F_{01} &=& E_1\ ,
\eeqa
\end{subequations}
where $B$ is the magnetic field perpendicular to the plane and $\mb{E}=(E_1,E_2)$ is the usual electric field in the plane. These 
formulas follow from the fact that when $\M=\mathbb{R}^2$ we can identify $A_0$ with the usual scalar potential in electromagnetism, 
while we have $A_j= -A^j$, where $A^j$ are the components of the usual vector potential $\mb{A}=(A^1,A^2)$. 
The usual electric and magnetic fields on flat space are then defined in terms of these by the usual formulas
\begin{subequations}
\beqa
	\mb{E} &=& -\mbs{\nabla} A_0 - \pd_0 \mb{A} \\
 	B &=& \pd_1 A^2 - \pd_2 A^1\ ,
\eeqa
\end{subequations}
where $\mbs{\nabla}=(\pd_1,\pd_2)$ is the ordinary gradient operator on flat space.


\begin{thebibliography}{36}%
\makeatletter
\providecommand \@ifxundefined [1]{%
 \@ifx{#1\undefined}
}%
\providecommand \@ifnum [1]{%
 \ifnum #1\expandafter \@firstoftwo
 \else \expandafter \@secondoftwo
 \fi
}%
\providecommand \@ifx [1]{%
 \ifx #1\expandafter \@firstoftwo
 \else \expandafter \@secondoftwo
 \fi
}%
\providecommand \natexlab [1]{#1}%
\providecommand \enquote  [1]{``#1''}%
\providecommand \bibnamefont  [1]{#1}%
\providecommand \bibfnamefont [1]{#1}%
\providecommand \citenamefont [1]{#1}%
\providecommand \href@noop [0]{\@secondoftwo}%
\providecommand \href [0]{\begingroup \@sanitize@url \@href}%
\providecommand \@href[1]{\@@startlink{#1}\@@href}%
\providecommand \@@href[1]{\endgroup#1\@@endlink}%
\providecommand \@sanitize@url [0]{\catcode `\\12\catcode `\$12\catcode
  `\&12\catcode `\#12\catcode `\^12\catcode `\_12\catcode `\%12\relax}%
\providecommand \@@startlink[1]{}%
\providecommand \@@endlink[0]{}%
\providecommand \url  [0]{\begingroup\@sanitize@url \@url }%
\providecommand \@url [1]{\endgroup\@href {#1}{\urlprefix }}%
\providecommand \urlprefix  [0]{URL }%
\providecommand \Eprint [0]{\href }%
\providecommand \doibase [0]{http://dx.doi.org/}%
\providecommand \selectlanguage [0]{\@gobble}%
\providecommand \bibinfo  [0]{\@secondoftwo}%
\providecommand \bibfield  [0]{\@secondoftwo}%
\providecommand \translation [1]{[#1]}%
\providecommand \BibitemOpen [0]{}%
\providecommand \bibitemStop [0]{}%
\providecommand \bibitemNoStop [0]{.\EOS\space}%
\providecommand \EOS [0]{\spacefactor3000\relax}%
\providecommand \BibitemShut  [1]{\csname bibitem#1\endcsname}%
\let\auto@bib@innerbib\@empty
\bibitem [{\citenamefont {Niemi}\ and\ \citenamefont {Semenoff}(1983)}]{NS}%
  \BibitemOpen
  \bibfield  {author} {\bibinfo {author} {\bibfnamefont {A.~J.}\ \bibnamefont
  {Niemi}}\ and\ \bibinfo {author} {\bibfnamefont {G.~W.}\ \bibnamefont
  {Semenoff}},\ }\href@noop {} {\bibfield  {journal} {\bibinfo  {journal}
  {Phys. Rev. Lett.}\ }\textbf {\bibinfo {volume} {51}},\ \bibinfo {pages}
  {2077} (\bibinfo {year} {1983})}\BibitemShut {NoStop}%
\bibitem [{\citenamefont {Redlich}(1984{\natexlab{a}})}]{redlich1}%
  \BibitemOpen
  \bibfield  {author} {\bibinfo {author} {\bibfnamefont {A.~N.}\ \bibnamefont
  {Redlich}},\ }\href@noop {} {\bibfield  {journal} {\bibinfo  {journal} {Phys.
  Rev. Lett.}\ }\textbf {\bibinfo {volume} {52}},\ \bibinfo {pages} {18}
  (\bibinfo {year} {1984}{\natexlab{a}})}\BibitemShut {NoStop}%
\bibitem [{\citenamefont {Redlich}(1984{\natexlab{b}})}]{redlich2}%
  \BibitemOpen
  \bibfield  {author} {\bibinfo {author} {\bibfnamefont {A.~N.}\ \bibnamefont
  {Redlich}},\ }\href@noop {} {\bibfield  {journal} {\bibinfo  {journal} {Phys.
  Rev. D}\ }\textbf {\bibinfo {volume} {29}},\ \bibinfo {pages} {2366}
  (\bibinfo {year} {1984}{\natexlab{b}})}\BibitemShut {NoStop}%
\bibitem [{\citenamefont {Alvarez-Gaume}\ \emph {et~al.}(1985)\citenamefont
  {Alvarez-Gaume}, \citenamefont {Della~Pietra},\ and\ \citenamefont
  {Moore}}]{anomalies-odd}%
  \BibitemOpen
  \bibfield  {author} {\bibinfo {author} {\bibfnamefont {L.}~\bibnamefont
  {Alvarez-Gaume}}, \bibinfo {author} {\bibfnamefont {S.}~\bibnamefont
  {Della~Pietra}}, \ and\ \bibinfo {author} {\bibfnamefont {G.}~\bibnamefont
  {Moore}},\ }\href@noop {} {\bibfield  {journal} {\bibinfo  {journal} {Ann.
  Phys.}\ }\textbf {\bibinfo {volume} {163}},\ \bibinfo {pages} {288} (\bibinfo
  {year} {1985})}\BibitemShut {NoStop}%
\bibitem [{\citenamefont {Witten}(2016{\natexlab{a}})}]{witten-parity}%
  \BibitemOpen
  \bibfield  {author} {\bibinfo {author} {\bibfnamefont {E.}~\bibnamefont
  {Witten}},\ }\href@noop {} {\bibfield  {journal} {\bibinfo  {journal} {Phys.
  Rev. B}\ }\textbf {\bibinfo {volume} {94}},\ \bibinfo {pages} {195150}
  (\bibinfo {year} {2016}{\natexlab{a}})}\BibitemShut {NoStop}%
\bibitem [{\citenamefont {Fu}\ \emph {et~al.}(2007)\citenamefont {Fu},
  \citenamefont {Kane},\ and\ \citenamefont {Mele}}]{fu-kane-mele}%
  \BibitemOpen
  \bibfield  {author} {\bibinfo {author} {\bibfnamefont {L.}~\bibnamefont
  {Fu}}, \bibinfo {author} {\bibfnamefont {C.~L.}\ \bibnamefont {Kane}}, \ and\
  \bibinfo {author} {\bibfnamefont {E.~J.}\ \bibnamefont {Mele}},\ }\href@noop
  {} {\bibfield  {journal} {\bibinfo  {journal} {Phys. Rev. Lett.}\ }\textbf
  {\bibinfo {volume} {98}},\ \bibinfo {pages} {106803} (\bibinfo {year}
  {2007})}\BibitemShut {NoStop}%
\bibitem [{\citenamefont {Qi}\ \emph {et~al.}(2008)\citenamefont {Qi},
  \citenamefont {Hughes},\ and\ \citenamefont {Zhang}}]{QHZ}%
  \BibitemOpen
  \bibfield  {author} {\bibinfo {author} {\bibfnamefont {X.-L.}\ \bibnamefont
  {Qi}}, \bibinfo {author} {\bibfnamefont {T.~L.}\ \bibnamefont {Hughes}}, \
  and\ \bibinfo {author} {\bibfnamefont {S.-C.}\ \bibnamefont {Zhang}},\
  }\href@noop {} {\bibfield  {journal} {\bibinfo  {journal} {Phys. Rev. B}\
  }\textbf {\bibinfo {volume} {78}},\ \bibinfo {pages} {195424} (\bibinfo
  {year} {2008})}\BibitemShut {NoStop}%
\bibitem [{\citenamefont {`t~Hooft}(1980)}]{hooft1980naturalness}%
  \BibitemOpen
  \bibfield  {author} {\bibinfo {author} {\bibfnamefont {G.}~\bibnamefont
  {`t~Hooft}},\ }\href@noop {} {\bibfield  {journal} {\bibinfo  {journal}
  {Recent Developments in Gauge Theories}\ ,\ \bibinfo {pages} {135}} (\bibinfo
  {year} {1980})}\BibitemShut {NoStop}%
\bibitem [{\citenamefont {Wen}(2013)}]{wen2013classifying}%
  \BibitemOpen
  \bibfield  {author} {\bibinfo {author} {\bibfnamefont {X.-G.}\ \bibnamefont
  {Wen}},\ }\href@noop {} {\bibfield  {journal} {\bibinfo  {journal} {Phys.
  Rev. D}\ }\textbf {\bibinfo {volume} {88}},\ \bibinfo {pages} {045013}
  (\bibinfo {year} {2013})}\BibitemShut {NoStop}%
\bibitem [{\citenamefont {Kapustin}(2014)}]{kapustin2014symmetry}%
  \BibitemOpen
  \bibfield  {author} {\bibinfo {author} {\bibfnamefont {A.}~\bibnamefont
  {Kapustin}},\ }\href@noop {} {\bibfield  {journal} {\bibinfo  {journal}
  {arXiv preprint arXiv:1403.1467}\ } (\bibinfo {year} {2014})}\BibitemShut
  {NoStop}%
\bibitem [{\citenamefont {Kapustin}\ and\ \citenamefont
  {Thorngren}(2014{\natexlab{a}})}]{kapustin2014anomalies}%
  \BibitemOpen
  \bibfield  {author} {\bibinfo {author} {\bibfnamefont {A.}~\bibnamefont
  {Kapustin}}\ and\ \bibinfo {author} {\bibfnamefont {R.}~\bibnamefont
  {Thorngren}},\ }\href@noop {} {\bibfield  {journal} {\bibinfo  {journal}
  {arXiv preprint arXiv:1404.3230}\ } (\bibinfo {year}
  {2014}{\natexlab{a}})}\BibitemShut {NoStop}%
\bibitem [{\citenamefont {Kapustin}\ and\ \citenamefont
  {Thorngren}(2014{\natexlab{b}})}]{kapustin2014anomalous}%
  \BibitemOpen
  \bibfield  {author} {\bibinfo {author} {\bibfnamefont {A.}~\bibnamefont
  {Kapustin}}\ and\ \bibinfo {author} {\bibfnamefont {R.}~\bibnamefont
  {Thorngren}},\ }\href@noop {} {\bibfield  {journal} {\bibinfo  {journal}
  {Phys. Rev. Lett.}\ }\textbf {\bibinfo {volume} {112}},\ \bibinfo {pages}
  {231602} (\bibinfo {year} {2014}{\natexlab{b}})}\BibitemShut {NoStop}%
\bibitem [{\citenamefont {M{\"u}ller}\ and\ \citenamefont
  {Szabo}(2018)}]{muller-szabo}%
  \BibitemOpen
  \bibfield  {author} {\bibinfo {author} {\bibfnamefont {L.}~\bibnamefont
  {M{\"u}ller}}\ and\ \bibinfo {author} {\bibfnamefont {R.~J.}\ \bibnamefont
  {Szabo}},\ }\href@noop {} {\bibfield  {journal} {\bibinfo  {journal} {Commun.
  Math. Phys.}\ }\textbf {\bibinfo {volume} {362}},\ \bibinfo {pages} {1049}
  (\bibinfo {year} {2018})}\BibitemShut {NoStop}%
\bibitem [{\citenamefont {Mulligan}\ and\ \citenamefont
  {Burnell}(2013)}]{mulligan-burnell}%
  \BibitemOpen
  \bibfield  {author} {\bibinfo {author} {\bibfnamefont {M.}~\bibnamefont
  {Mulligan}}\ and\ \bibinfo {author} {\bibfnamefont {F.~J.}\ \bibnamefont
  {Burnell}},\ }\href@noop {} {\bibfield  {journal} {\bibinfo  {journal} {Phys.
  Rev. B}\ }\textbf {\bibinfo {volume} {88}},\ \bibinfo {pages} {085104}
  (\bibinfo {year} {2013})}\BibitemShut {NoStop}%
\bibitem [{\citenamefont {Witten}(2016{\natexlab{b}})}]{witten-fermions}%
  \BibitemOpen
  \bibfield  {author} {\bibinfo {author} {\bibfnamefont {E.}~\bibnamefont
  {Witten}},\ }\href@noop {} {\bibfield  {journal} {\bibinfo  {journal} {Rev.
  Mod. Phys.}\ }\textbf {\bibinfo {volume} {88}},\ \bibinfo {pages} {035001}
  (\bibinfo {year} {2016}{\natexlab{b}})}\BibitemShut {NoStop}%
\bibitem [{\citenamefont {Seiberg}\ and\ \citenamefont
  {Witten}(2016)}]{seiberg-witten}%
  \BibitemOpen
  \bibfield  {author} {\bibinfo {author} {\bibfnamefont {N.}~\bibnamefont
  {Seiberg}}\ and\ \bibinfo {author} {\bibfnamefont {E.}~\bibnamefont
  {Witten}},\ }\href@noop {} {\bibfield  {journal} {\bibinfo  {journal} {Progr.
  Theor. Exp. Phys.}\ }\textbf {\bibinfo {volume} {2016}} (\bibinfo {year}
  {2016})}\BibitemShut {NoStop}%
\bibitem [{\citenamefont {Kurkov}\ and\ \citenamefont
  {Vassilevich}(2017)}]{PhysRevD.96.025011}%
  \BibitemOpen
  \bibfield  {author} {\bibinfo {author} {\bibfnamefont {M.}~\bibnamefont
  {Kurkov}}\ and\ \bibinfo {author} {\bibfnamefont {D.}~\bibnamefont
  {Vassilevich}},\ }\href@noop {} {\bibfield  {journal} {\bibinfo  {journal}
  {Phys. Rev. D}\ }\textbf {\bibinfo {volume} {96}},\ \bibinfo {pages} {025011}
  (\bibinfo {year} {2017})}\BibitemShut {NoStop}%
\bibitem [{\citenamefont {Kurkov}\ and\ \citenamefont
  {Vassilevich}(2018)}]{kurkov2018gravitational}%
  \BibitemOpen
  \bibfield  {author} {\bibinfo {author} {\bibfnamefont {M.}~\bibnamefont
  {Kurkov}}\ and\ \bibinfo {author} {\bibfnamefont {D.}~\bibnamefont
  {Vassilevich}},\ }\href@noop {} {\bibfield  {journal} {\bibinfo  {journal}
  {J. High Energy Phys.}\ }\textbf {\bibinfo {volume} {2018}},\ \bibinfo
  {pages} {72} (\bibinfo {year} {2018})}\BibitemShut {NoStop}%
\bibitem [{\citenamefont {Hsieh}\ \emph {et~al.}(2008)\citenamefont {Hsieh},
  \citenamefont {Qian}, \citenamefont {Wray}, \citenamefont {Xia},
  \citenamefont {Hor}, \citenamefont {Cava},\ and\ \citenamefont
  {Hasan}}]{hsieh2008topological}%
  \BibitemOpen
  \bibfield  {author} {\bibinfo {author} {\bibfnamefont {D.}~\bibnamefont
  {Hsieh}}, \bibinfo {author} {\bibfnamefont {D.}~\bibnamefont {Qian}},
  \bibinfo {author} {\bibfnamefont {L.}~\bibnamefont {Wray}}, \bibinfo {author}
  {\bibfnamefont {Y.}~\bibnamefont {Xia}}, \bibinfo {author} {\bibfnamefont
  {Y.~S.}\ \bibnamefont {Hor}}, \bibinfo {author} {\bibfnamefont {R.~J.}\
  \bibnamefont {Cava}}, \ and\ \bibinfo {author} {\bibfnamefont {M.~Z.}\
  \bibnamefont {Hasan}},\ }\href@noop {} {\bibfield  {journal} {\bibinfo
  {journal} {Nature}\ }\textbf {\bibinfo {volume} {452}},\ \bibinfo {pages}
  {970} (\bibinfo {year} {2008})}\BibitemShut {NoStop}%
\bibitem [{\citenamefont {Hasan}\ and\ \citenamefont
  {Kane}(2010)}]{hasan-kane}%
  \BibitemOpen
  \bibfield  {author} {\bibinfo {author} {\bibfnamefont {M.~Z.}\ \bibnamefont
  {Hasan}}\ and\ \bibinfo {author} {\bibfnamefont {C.~L.}\ \bibnamefont
  {Kane}},\ }\href@noop {} {\bibfield  {journal} {\bibinfo  {journal} {Rev.
  Mod. Phys.}\ }\textbf {\bibinfo {volume} {82}},\ \bibinfo {pages} {3045}
  (\bibinfo {year} {2010})}\BibitemShut {NoStop}%
\bibitem [{\citenamefont {Atiyah}\ \emph {et~al.}(1975)\citenamefont {Atiyah},
  \citenamefont {Patodi},\ and\ \citenamefont {Singer}}]{APS1}%
  \BibitemOpen
  \bibfield  {author} {\bibinfo {author} {\bibfnamefont {M.~F.}\ \bibnamefont
  {Atiyah}}, \bibinfo {author} {\bibfnamefont {V.~K.}\ \bibnamefont {Patodi}},
  \ and\ \bibinfo {author} {\bibfnamefont {I.~M.}\ \bibnamefont {Singer}},\
  }in\ \href@noop {} {\emph {\bibinfo {booktitle} {Mathematical Proceedings of
  the Cambridge Philosophical Society}}},\ Vol.~\bibinfo {volume} {77}\
  (\bibinfo {organization} {Cambridge Univ Press},\ \bibinfo {year} {1975})\
  pp.\ \bibinfo {pages} {43--69}\BibitemShut {NoStop}%
\bibitem [{\citenamefont {Jackiw}(1984)}]{jackiw1984}%
  \BibitemOpen
  \bibfield  {author} {\bibinfo {author} {\bibfnamefont {R.}~\bibnamefont
  {Jackiw}},\ }\href@noop {} {\bibfield  {journal} {\bibinfo  {journal} {Phys.
  Rev. D}\ }\textbf {\bibinfo {volume} {29}},\ \bibinfo {pages} {2375}
  (\bibinfo {year} {1984})}\BibitemShut {NoStop}%
\bibitem [{\citenamefont {Barci}\ \emph {et~al.}(2001)\citenamefont {Barci},
  \citenamefont {Neto}, \citenamefont {Oxman},\ and\ \citenamefont
  {Sorella}}]{barci2001point}%
  \BibitemOpen
  \bibfield  {author} {\bibinfo {author} {\bibfnamefont {D.~G.}\ \bibnamefont
  {Barci}}, \bibinfo {author} {\bibfnamefont {J.~F.~M.}\ \bibnamefont {Neto}},
  \bibinfo {author} {\bibfnamefont {L.~E.}\ \bibnamefont {Oxman}}, \ and\
  \bibinfo {author} {\bibfnamefont {S.~P.}\ \bibnamefont {Sorella}},\
  }\href@noop {} {\bibfield  {journal} {\bibinfo  {journal} {Nucl. Phys. B}\
  }\textbf {\bibinfo {volume} {600}},\ \bibinfo {pages} {203} (\bibinfo {year}
  {2001})}\BibitemShut {NoStop}%
\bibitem [{\citenamefont {Haldane}(1988)}]{haldane-parity}%
  \BibitemOpen
  \bibfield  {author} {\bibinfo {author} {\bibfnamefont {F.~D.~M.}\
  \bibnamefont {Haldane}},\ }\href@noop {} {\bibfield  {journal} {\bibinfo
  {journal} {Phys. Rev. Lett.}\ }\textbf {\bibinfo {volume} {61}},\ \bibinfo
  {pages} {2015} (\bibinfo {year} {1988})}\BibitemShut {NoStop}%
\bibitem [{\citenamefont {Paranjape}\ and\ \citenamefont
  {Semenoff}(1983)}]{PS}%
  \BibitemOpen
  \bibfield  {author} {\bibinfo {author} {\bibfnamefont {M.~B.}\ \bibnamefont
  {Paranjape}}\ and\ \bibinfo {author} {\bibfnamefont {G.~W.}\ \bibnamefont
  {Semenoff}},\ }\href@noop {} {\bibfield  {journal} {\bibinfo  {journal}
  {Phys. Lett. B}\ }\textbf {\bibinfo {volume} {132}},\ \bibinfo {pages} {369}
  (\bibinfo {year} {1983})}\BibitemShut {NoStop}%
\bibitem [{\citenamefont {Atiyah}\ and\ \citenamefont {Singer}(1968)}]{AS}%
  \BibitemOpen
  \bibfield  {author} {\bibinfo {author} {\bibfnamefont {M.~F.}\ \bibnamefont
  {Atiyah}}\ and\ \bibinfo {author} {\bibfnamefont {I.~M.}\ \bibnamefont
  {Singer}},\ }\href@noop {} {\bibfield  {journal} {\bibinfo  {journal} {Ann.
  Math.}\ ,\ \bibinfo {pages} {546}} (\bibinfo {year} {1968})}\BibitemShut
  {NoStop}%
\bibitem [{\citenamefont {Eguchi}\ \emph {et~al.}(1980)\citenamefont {Eguchi},
  \citenamefont {Gilkey},\ and\ \citenamefont {Hanson}}]{EGH}%
  \BibitemOpen
  \bibfield  {author} {\bibinfo {author} {\bibfnamefont {T.}~\bibnamefont
  {Eguchi}}, \bibinfo {author} {\bibfnamefont {P.~B.}\ \bibnamefont {Gilkey}},
  \ and\ \bibinfo {author} {\bibfnamefont {A.~J.}\ \bibnamefont {Hanson}},\
  }\href@noop {} {\bibfield  {journal} {\bibinfo  {journal} {Phys. Rep.}\
  }\textbf {\bibinfo {volume} {66}},\ \bibinfo {pages} {213} (\bibinfo {year}
  {1980})}\BibitemShut {NoStop}%
\bibitem [{\citenamefont {Thouless}\ \emph {et~al.}(1982)\citenamefont
  {Thouless}, \citenamefont {Kohmoto}, \citenamefont {Nightingale},\ and\
  \citenamefont {den Nijs}}]{thouless1982quantized}%
  \BibitemOpen
  \bibfield  {author} {\bibinfo {author} {\bibfnamefont {D.~J.}\ \bibnamefont
  {Thouless}}, \bibinfo {author} {\bibfnamefont {M.}~\bibnamefont {Kohmoto}},
  \bibinfo {author} {\bibfnamefont {M.~P.}\ \bibnamefont {Nightingale}}, \ and\
  \bibinfo {author} {\bibfnamefont {M.}~\bibnamefont {den Nijs}},\ }\href@noop
  {} {\bibfield  {journal} {\bibinfo  {journal} {Phys. Rev. Lett.}\ }\textbf
  {\bibinfo {volume} {49}},\ \bibinfo {pages} {405} (\bibinfo {year}
  {1982})}\BibitemShut {NoStop}%
\bibitem [{\citenamefont {Xiao}\ \emph {et~al.}(2010)\citenamefont {Xiao},
  \citenamefont {Chang},\ and\ \citenamefont {Niu}}]{niu-review}%
  \BibitemOpen
  \bibfield  {author} {\bibinfo {author} {\bibfnamefont {D.}~\bibnamefont
  {Xiao}}, \bibinfo {author} {\bibfnamefont {M.-C.}\ \bibnamefont {Chang}}, \
  and\ \bibinfo {author} {\bibfnamefont {Q.}~\bibnamefont {Niu}},\ }\href@noop
  {} {\bibfield  {journal} {\bibinfo  {journal} {Rev. Mod. Phys.}\ }\textbf
  {\bibinfo {volume} {82}},\ \bibinfo {pages} {1959} (\bibinfo {year}
  {2010})}\BibitemShut {NoStop}%
\bibitem [{\citenamefont {Vishwanath}\ and\ \citenamefont
  {Senthil}(2013)}]{VS2013}%
  \BibitemOpen
  \bibfield  {author} {\bibinfo {author} {\bibfnamefont {A.}~\bibnamefont
  {Vishwanath}}\ and\ \bibinfo {author} {\bibfnamefont {T.}~\bibnamefont
  {Senthil}},\ }\href@noop {} {\bibfield  {journal} {\bibinfo  {journal} {Phys.
  Rev. X}\ }\textbf {\bibinfo {volume} {3}},\ \bibinfo {pages} {011016}
  (\bibinfo {year} {2013})}\BibitemShut {NoStop}%
\bibitem [{\citenamefont {Metlitski}\ \emph {et~al.}(2013)\citenamefont
  {Metlitski}, \citenamefont {Kane},\ and\ \citenamefont {Fisher}}]{MKF1}%
  \BibitemOpen
  \bibfield  {author} {\bibinfo {author} {\bibfnamefont {M.~A.}\ \bibnamefont
  {Metlitski}}, \bibinfo {author} {\bibfnamefont {C.~L.}\ \bibnamefont {Kane}},
  \ and\ \bibinfo {author} {\bibfnamefont {M.~P.~A.}\ \bibnamefont {Fisher}},\
  }\href@noop {} {\bibfield  {journal} {\bibinfo  {journal} {Phys. Rev. B}\
  }\textbf {\bibinfo {volume} {88}},\ \bibinfo {pages} {035131} (\bibinfo
  {year} {2013})}\BibitemShut {NoStop}%
\bibitem [{\citenamefont {Lapa}\ \emph {et~al.}(2017)\citenamefont {Lapa},
  \citenamefont {Jian}, \citenamefont {Ye},\ and\ \citenamefont
  {Hughes}}]{lapa1}%
  \BibitemOpen
  \bibfield  {author} {\bibinfo {author} {\bibfnamefont {M.~F.}\ \bibnamefont
  {Lapa}}, \bibinfo {author} {\bibfnamefont {C.-M.}\ \bibnamefont {Jian}},
  \bibinfo {author} {\bibfnamefont {P.}~\bibnamefont {Ye}}, \ and\ \bibinfo
  {author} {\bibfnamefont {T.~L.}\ \bibnamefont {Hughes}},\ }\href@noop {}
  {\bibfield  {journal} {\bibinfo  {journal} {Phys. Rev. B}\ }\textbf {\bibinfo
  {volume} {95}},\ \bibinfo {pages} {035149} (\bibinfo {year}
  {2017})}\BibitemShut {NoStop}%
\bibitem [{\citenamefont {Lu}\ and\ \citenamefont
  {Vishwanath}(2012)}]{lu-vishwanath}%
  \BibitemOpen
  \bibfield  {author} {\bibinfo {author} {\bibfnamefont {Y.-M.}\ \bibnamefont
  {Lu}}\ and\ \bibinfo {author} {\bibfnamefont {A.}~\bibnamefont
  {Vishwanath}},\ }\href@noop {} {\bibfield  {journal} {\bibinfo  {journal}
  {Phys. Rev. B}\ }\textbf {\bibinfo {volume} {86}},\ \bibinfo {pages} {125119}
  (\bibinfo {year} {2012})}\BibitemShut {NoStop}%
\bibitem [{\citenamefont {Senthil}\ and\ \citenamefont
  {Levin}(2013)}]{levin-senthil}%
  \BibitemOpen
  \bibfield  {author} {\bibinfo {author} {\bibfnamefont {T.}~\bibnamefont
  {Senthil}}\ and\ \bibinfo {author} {\bibfnamefont {M.}~\bibnamefont
  {Levin}},\ }\href@noop {} {\bibfield  {journal} {\bibinfo  {journal} {Phys.
  Rev. Lett.}\ }\textbf {\bibinfo {volume} {110}},\ \bibinfo {pages} {046801}
  (\bibinfo {year} {2013})}\BibitemShut {NoStop}%
\bibitem [{\citenamefont {Lapa}\ and\ \citenamefont {Hughes}(2017)}]{lapa2}%
  \BibitemOpen
  \bibfield  {author} {\bibinfo {author} {\bibfnamefont {M.~F.}\ \bibnamefont
  {Lapa}}\ and\ \bibinfo {author} {\bibfnamefont {T.~L.}\ \bibnamefont
  {Hughes}},\ }\href@noop {} {\bibfield  {journal} {\bibinfo  {journal} {Phys.
  Rev. B}\ }\textbf {\bibinfo {volume} {96}},\ \bibinfo {pages} {115123}
  (\bibinfo {year} {2017})}\BibitemShut {NoStop}%
\bibitem [{\citenamefont {Elitzur}\ \emph {et~al.}(1986)\citenamefont
  {Elitzur}, \citenamefont {Rabinovici}, \citenamefont {Frishman},\ and\
  \citenamefont {Schwimmer}}]{elitzur1986origins}%
  \BibitemOpen
  \bibfield  {author} {\bibinfo {author} {\bibfnamefont {S.}~\bibnamefont
  {Elitzur}}, \bibinfo {author} {\bibfnamefont {E.}~\bibnamefont {Rabinovici}},
  \bibinfo {author} {\bibfnamefont {Y.}~\bibnamefont {Frishman}}, \ and\
  \bibinfo {author} {\bibfnamefont {A.}~\bibnamefont {Schwimmer}},\ }\href@noop
  {} {\bibfield  {journal} {\bibinfo  {journal} {Nucl. Phys. B}\ }\textbf
  {\bibinfo {volume} {273}},\ \bibinfo {pages} {93} (\bibinfo {year}
  {1986})}\BibitemShut {NoStop}%
\end{thebibliography}

%

\end{document}